# Surface electronic properties of diamond

C.E. Nebel

Fraunhofer-Institute for Applied Solid State Physics, Tullastrasse 72, 79108 Freiburg, Germany

**Abstract**

Surface electronic properties of undoped hydrogen terminated diamond covered with adsorbates or in electrolyte solutions are summarized. The formation of a conductive layer at the surface of diamond is discussed based on Hall effect, conductivity, contact potential difference (CPM), scanning electron microscopy (SEM), and cyclic voltammetry data applied on homoepitaxially grown CVD diamond films with atomically smooth hydrogen terminated surfaces. Due to electron transfer from valence band states into empty states of the electrolyte ("transfer doping"), a highly conductive surface layer is generated. Holes propagate in the layer with mobilities up to 350 cm$^2$/Vs. The sheet hole density is in the range $10^{11}$ to $5 \times 10^{12}$ cm$^{-2}$, and dependents on pH of the electrolyte. Numerical solutions of the Schrödinger and Poisson equations reveal a 2D density of state (DOS) distribution. This has been utilized to manufacture ion-sensitive field effect transistors (ISFET). The drain source conductivity is pH dependent, with about 66 mV/pH. Application of potentials larger than the oxidation threshold of +0.7 V (pH 13) to +1.6 V (pH 1) gives rise to strong leakage currents and to partial surface oxidation.

# Inhalt

# 1. Introduction

Diamond exhibits considerable potential for a variety of applications on account of its outstanding mechanical, optical and electronic properties (Dischler, B., Wild, C. 1998). Especially, two features of diamond are unique among all semiconductors (Bandis, C., Pate, B. 1995; Ristein, J., Stein, W., Ley, L. 1997). One is the true negative electron affinity (NEA) of the hydrogen terminated diamond surface. Secondly, diamond shows under special circumstances a high p-type surface conductivity (Landstrass, M.I., Ravi, K.V. 1989a; Landstrass, M.I., Ravi, K.V. 1989b) which may be utilized in a variety of electronic devices (Geis M.W., Twichell, J.C. 1995; Gluche, P., Aleksov, A., Ebert, W., Kohn, E. 1997; . Tsugawa, K., Kitatani, K., Noda, H. et al 1999). The electron affinity $\chi$ is the energy difference between the conduction-band minimum $E_{CBM}$ and the vacuum level $E_{VAC}$ (Mönch, W., 1995):

$$\chi = E_{VAC} - E_{CBM}$$

It can be modified by surface dipole effects which are intentionally or unintentionally generated on the surface of diamond. Diamond therefore shows great potential for cold electron emitters (Geis M.W., Twichell, J.C. 1995), optimized molecular interfaces (Nebel, C.E. 2007; Hoffmann, R., Kriele, A., Obloh, H. et al., 2011), pH- and bio-sensors due to the combination of surface conductivity, biocompatibility, and chemical stability (Nebel, C.E., Rezek, B., Shin, D., Uetsuka, H., Yang, N. 2007).

This chapter is organized as follows: Firstly, we will introduce the effects of dipoles on diamond surfaces (part II), then we will discuss the surface conductivity of undoped diamond in air (part III), followed by a description of surface electronic properties of diamond covered with adsorbates (part IV). A summary of features related to surface electronic properties in electrolytes will be given in part V and a summary in part VI.

# 2. Dipole Effects on diamond surfaces

The diamond surface can be terminated by a variety of elements like H, O, F and CL (Miller, J.B., Brown, D.W. 1996) or organic molecules like OH, phenyl, long-chain amines etc. (Nebel, C.E., Rezek, B., Shin, D., Uetsuka, H., Yang, N. 2007, Nebel, C.E., Shin, D., Nakamura, T., et al 2007), giving rise to new phenomena and applications. Surface terminations affect the electron affinity $\chi$. For example, oxygen (O) termination causes an increase of the positive electron affinities (PEA), while hydrogen (H) gives rise to a negative electron affinity (NEA) (Bandis, C., Pate, B.B. 1995; Ristein, J., Maier, F., Riedel, M., Cui, J.B., Ley, L. 2000; Maier, F., Ristein, J., Ley, L. 2001; Cui, J.B., Ristein, J., Ley, L., 1998). This is shown schematically in Fig. 1. The variation of the electron affinity can be discussed taking into account the electronegativity, $\delta$, of atoms (Mönch, W., 1995). $\delta$ is a chemical property that describes the tendency of an atom to attract electrons towards itself. The higher the associated electronegativity number, the more an atom attracts electrons. The electronegativity of atoms has been calculated and is given in a dimensionless quantity, the Pauling units (Pauling, L., 1939/1960). Carbon has an electronegativity of 2.55 and hydrogen 2.2. Pauling correlated the amount of ionic character or the iconicity of single bonds in diatomic molecules **A - B** with the different $\delta_A$ - $\delta_B$ of the atomic electronegativities of the atoms forming the molecule. A revised version of the relation originally proposed by Pauling is that of Hanney and Smith (Hanney, N.B., Smith, C.P., 1946):

$$\Delta q_1 = 0.16 \, |\delta_A - \delta_B| + 0.035 \, |\delta_A - \delta_B|^2 \quad \text{(Eq. 1)}$$

In a simple point-charge model, such atoms are charge by $+\Delta q_1 e$ and $-\Delta q_1 e$ where the more electronegative atom becomes negatively charged ("e" is the elementary charge). Diatomic molecules with $|\delta_A - \delta_B| \neq 0$ thus possess dipole moments. By using the simple point-charge model, they may be written as:

$$p = \Delta q_1 e d \quad \text{(Eq. 2)}$$

where *d* is the bond length of the diatomic molecule.

Considering the hydrogen terminated diamond surface diatomic surface dipoles $^-$**C - H**$^+$ are formed. In case of oxygen termination (electronegativity of O is 3.44) the dipole reverses to $^+$**C – O**$^-$. Such dipoles cause potential steps *ΔV* perpendicular to the surface over the distance of the C-H bond length of $d_{CH}$ = 1.1 Å or over the spacing between C-O (ether configuration of O) of about $d_{CO}$ = 0.7 Å. This potential step gives rise to a lowering (H-termination) or enlargement (O-termination) of the vacuum level, $E_{VAC}$, with respect the conduction-band minimum, $E_{CM}$, given by *ΔE = eΔV* compared to its value without the dipole layer.

The reduction of the electron affinity $\chi$ from $\chi_{max}$ depends on the areal density, *n*, and on the magnitude *p* of the dipoles (Maier, F., Ristein, J., Ley, L. 2001; Cui, J.B., Ristein, J., Ley, L., 1998; Kawarada, H. 1996):

$$\chi - \chi_{max} = -e\Delta V = -\frac{epn}{\varepsilon_o} f(n) \quad \text{(Eq. 3)}$$

where $\varepsilon_o$ is the dielectric constant. The function *f(n)* which depends on n takes the interaction of dipoles into account with the result that the contribution of each dipole to *ΔV* is reduced for high dipole densities. An expression for f(n) with the polarizability $\alpha$ of the dipoles as a parameter can be obtained according to the calculation of Topping (Topping, J. 1927):

$$f(n) = \left(1 + \frac{9\alpha n^{3/2}}{4\pi\varepsilon_o}\right)^{-1} \quad \text{(Eq. 4)}$$

Maier, F., Ristein, J., Ley, L. 2001; and Cui, J.B., Ristein, J., Ley, L., 1998 have discussed these properties in detail and their results are summarized in Table 1:

| Diamond surface | Electron affinity $\chi$ | Dipole moment p |
|---|---|---|
| | Unit: *eV* | Unit: *e Å* |
| (111)-(2x1) | + 0.38 | |
| (111)-(2x1):H | - 1.27 | + 0.09 |
| (100)-(2x1) | + 0.50 | |
| (100)-(2x1):H | - 1.30 | + 0.08 |
| (100)-(1x1):O | + 1.73 | -0.10 |

Table I (data from Maier, F., Ristein, J., Ley, L. 2001; and Cui, J.B., Ristein, J., Ley, L., 1998).

Instead of using fully hydrogen or oxygen terminated diamond surfaces with data summarized in Table I there is the possibility to apply partially terminated surfaces generated for example by thermal evaporation of hydrogen or oxygen. Considering first order desorption kinetics of hydrogen from the diamond surface, the dipole density as a function of isothermal annealing time t is given by (Maier, F., Ristein, J., Ley, L. 2001; and Cui, J.B., Ristein, J., Ley, L., 1989):

$n(t) = n_o e^{-t/\tau}$    (Eq. 5)

where $n_o$ is the areal density of H atoms at the start of the annealing and equals thus the surface density of C atoms of $1.81 \times 10^{15}$ cm$^{-2}$ for the (111) surface and $1.57 \times 10^{15}$ cm$^{-2}$ on (100). For thermal desorption experiments on H-terminated diamond at 1000 K τ is about 1850 s. A typical result of the variation of the electron affinity as measured by photoemission experiments on H-terminated diamond is shown in Fig. 2. The electron affinity χ versus annealing time (at T= 1000 K) is continuously changing from -1.3 eV (fully hydrogen terminated) to +0.3 eV for a bare carbon surface. The dashed line through the date points is a result of a fit calculation using Eq. 3, 4 and 5 (for details see Maier, F., Ristein, J., Ley, L. 2001 and Cui, J.B., Ristein, J., Ley, L., 1998). To investigate the effect of increasing positive electron affinity thermal annealing experiments in combination with photoemission measurements and XPS have been performed on an initially oxidized diamond surface with (100)-orientation. With decreasing oxygen coverage, the electron affinity decreases from +1.7 eV to 0.3 eV for a bare diamond surface as shown in Fig. 3.

These experiments demonstrate that the electron affinity on diamond can be tuned between -1.3 eV and +1.7 eV over about 3 eV using fractional H- and O-terminations. It means that the vacuum energy level can be adjusted by surface treatments thereby adjusting energy levels (HOMO, LUMO) of molecular layers on diamond for optimization of electron transfer rates in electrochemical devices or at biomimetic diamond/protein interfaces.

## 3. Surface conductivity of undoped diamond in air

Insulating ("undoped") hydrogen terminated diamond shows in air a surface conductivity (Maier, F., Riedel, M., Mantel, B., Ristein, J., Ley L., 2000, Gi, G.S., Mizumasa, T., Akiba, Y., et al., 1995). Hall effect experiments revealed the p-type nature of the surface conductivity (Maki, T.,Shikama, S., Komori, M., et al. 1992, Hayashi, K., Yamanaka, S., Okushi, H., Kajimura, K., 1996), with typical hole sheet-densities between ($10^{10} – 10^{13}$) cm$^{-2}$, and Hall mobilities between 1 and 100 cm$^2$/Vs (Looi, H.J., Jackman, R.B., Foord, J.S., 1998, Gi, R.S., Tashiro, K., Tanaka, S., et al. 1999). A summary of data is shown in Fig. 4 (Nebel, C.E., Rezek, B., Shin, D., Watanabe, H., 2006). There is an overall trend towards lower mobilities with increasing hole sheet-densities.

The surface conductivity is related to hydrogen termination in combination with an adsorbate coverage. As most plausible model for the formation of the surface conductivity a "transfer doping model" has been proposed (Gi, R.S., Mizumasa, T., Akiba, Y. et al. 1995, Shirafuji, J., Sugino, T., 1996, Maier, F., Riedel, M., Mantel, B., Ristein, J., Ley, L., 2000, Chakrapani, V., Angus, J.C., Anderson, A.B., et al. 2007). In this model, valence-band electrons tunnel into electronic empty states of an adjacent adsorbate layer as shown schematically in Fig. 5. In order to act as sink for electrons, the adsorbate layer must have its lowest unoccupied electronic level below the valence band maximum (VBM) of diamond. Maier et al. (Maier, F., Riedel, M., Mantel, B., Ristein, J., Ley, L., 2000) proposed that for standard atmospheric conditions, the pH value of water is about 6 due to $CO_2$ content or other ionic contaminations. They calculated the chemical potential $\mu_e$ for such an aqueous wetting layer to be about -4.26 eV below the vacuum level. To calculate the valence band maximum, $E_{VBM}$, of H-terminated diamond with respect to the vacuum level, $E_{VAC}$, we take into account the band gap of diamond, $E_G$ of 5.47 eV, and the negative electron affinity $\chi$ of -1.1 to -1.3 eV (Maier, F., Riedel, M., Mantel, B., Ristein, J., Ley, L. 2000, Takeuchi, D., Kato, H., Ri, G.S.I, et al. 2005). This results in a gap between vacuum level and valence band maximum of -4.17 to -4.37 eV if we refer to the vacuum level as zero. The Fermi-level $E_F$ at the electrolyte/diamond interface is therefore either slightly in (about 90 meV) or close to the valence band maximum (110 meV above $E_{VBM}$). A generalized summary of chemical potentials with respect to hydrogen and oxygen terminated diamond is given in Fig. 6 (Angus, J.C., Pleskov, Y.V., Eaton, S.C., 2004). It shows that for well-defined pH liquids the chemical potential $\mu(pH)$ may indeed be below $E_{VBM}$. In such a case a hole accumulation layer at the surface of diamond is generated by transfer doping. However, as the pH value of adsorbate layers cannot be detected experimentally, a discussion based on assumptions will not elucidate the real features.

In the following a summary of surface electronic properties of H-terminated diamond is given for a) diamond/adsorbate and b) diamond/electrolyte combinations based on Hall effect, contact potential difference ("Kelvin force"), Schottky junction characterizations, cyclic voltammetry, and field effect measurements using ion-sensitive field effect transistors (ISFET). In addition, the electronic properties of such interfaces have been calculated using one-dimensional numerical solutions of the Schrödinger and Poisson equations.

## 4. Surface electronic properties of diamond covered with adsorbates

### 4.1 Contact potential difference (CPD) experiments

For contact potential difference (CPD) experiments on diamond, Au is used as a reference metal (Rezek, B., Shin, D., Watanabe, H., Nebel, C.E. 2007). However, the work function $\chi$ of Au exposed to air as measured by total photo-yield experiments is typically 4.3 (± 0.1) eV which is significantly smaller than values reported in the literature for Au of 4.9 to 5.1 eV (Sze, S.M., 1981). A typical scanning electron microscopy image (SEM) and a related CPD result as detected on Au (region A), hydrogen terminated diamond surface (region B) and on a surface which has been exposed to oxygen plasma (regions C) are shown in Figs. 7a and 7b. In Fig. 7c a line scan in units of potential (mV) (white line in Fig. 7) is shown which show no difference between Au and H-terminated diamond. The oxidized area is dark in SEM, which is a result of the lower electron back scattering from the positive electron affinity surface. The hydrogen terminated surface appears bright due to its negative electron affinity. Figure 5 summarizes the result schematically, taking into account a negative electron affinity of -1.2 eV (± 0.1 eV) (Cui, J.B., Ristein, J., Ley, L. 1998, Nebel, C.E.; Rezek, B.; Zrenner, A., 2004a,b) and surface energies as detected with respect to the work-function $\chi_{Au}$ of Au.

CPD measurements on Al contacts on H-terminated diamond result in a surface potential difference of +588 mV. The work function of Al, $\chi_{Al}$, is therefore about 588 meV smaller than the work function of Au, which results in $\chi_{Al}$ = 3.7 eV. We assume that this result is governed by partially oxidation of the Al surface as $\chi_{Al}$ reported in the literature is 4.3 eV (Sze, S.M. 1981). Fig. 8 summarizes the results schematically. Based on CPD data, the Fermi level of H-terminated diamond covered with an adsorbate layer is slightly in the valence band (in this case, about 30 meV). Obviously, no band-bending is generated by Au-contacts on H-terminated diamond resulting in "Ohmic" contact properties. This is different for Al where a band bending of about 570 meV is expected as shown in Fig.9.

## 4.2 Current-voltage (IV) properties

The current voltage characteristics of Al on H-terminated diamond, as measured at 300 K, are shown in Fig. 10 (Al size: 250 $\mu$m x 250 $\mu$m) measured on two different junctions. They show Schottky properties with an ideality factor n of about 1. Applying negative voltages of more than 0.6 V (threshold voltage) to the Al contacts give rise to an exponential increase of current over 7 orders of magnitude. The threshold voltage of 600 meV is reasonably well in agreement with the detected energy barrier of 570 meV by CPD experiments. Positive voltages result in minor current variations (reverse currents) in the range of $10^{-13}$ A (note, the current has not been normalized by contact areas, as it is an in-plane current, flowing at the surface of diamond between Al and Au).

## 4.3 Capacitance-voltage (CV) experiments

Capacitance-voltage (CV) experiments on these contacts measured at T = 300 K are shown in Fig. 11 (Garrido, J.A., Nebel, C.E., Stutzmann, M., Snidero, E., Bergonzo, P., 2002). It is important to note that the capacitance, C, is ≤ 1 pF in the regime -3 Volt ≤ V ≤ 0 Volt. If a three-dimensional parallel-plate capacitor is assumed, the capacitance C can be calculated by:

*C= $\varepsilon_o$ $\varepsilon_r$(A/d)*     (Eq. 6)

where $\varepsilon_o$ is the dielectric constant, $\varepsilon_r$ is the relative dielectric constant of diamond ($\varepsilon_r$ = 5.7), A is the area of the Al contact, and d is the depletion layer width or the distance between the Al top contact and the doping layer. For an Al contact with 250 x 250 $\mu m^2$ area and a spatial distance of 10 nm between the Al contact and the H-induced doping layer, the capacitance would be about 310 pF. Experimentally, C is ≤1 pF on all contacts realized and follows an approximate linear relationship with the length of the periphery of the Al contact as shown in Fig. 12. This indicates that no hole accumulation layer is present below the Al-layer. The diode characteristics are governed by in-plane properties, where a p-type channel exists at the surface of hydrogen terminated diamond which connects to Al or Au. Unfortunately, 2D- and 3D Schottky junctions are generating nearly comparable IV and CV properties as discussed in detail by Petrosyan and Shik (Petrosyan, S.G., Shik, A. Y., 1989) and by Gelmont and Shur (Gelmont, B., Shur M., 1992).

## 4.4 Two dimensional properties of a perfectly H-terminated diamond surface

Figure 5 shows a schematic view of the electronic properties at the surface of H-terminated diamond, where valence-band electrons can tunnel into the empty electronic states of an adsorbate layer. Tunneling gives rise to band-bending which decreases in diamond with increasing distance to the surface. To calculate the width and bending of the valence band, one has to take into account light hole (LL), heavy hole (HH) and split-off (SO) bands. The band structure of diamond is described by Luttinger parameters and has been discussed by Willatzen, Cardona and Christensen (Willatzen, M., Cardona, M., Christensen, N.E., 1994). They derive $\gamma_1$ = 2.54, $\gamma_2$ = -0.1, and $\gamma_3$ = 0.63.

To calculate the band-bending in the vicinity of the surface of diamond a numeric approach has been used, optimized for solving the Schrödinger and Poisson equations simultaneously to calculate the energy levels and the electron wave-functions in narrow GaAs/Ga$_{1-x}$Al$_x$As hetero-junctions. Details can be found in Stern and DasSarma (Stern, F., DasSarma S. 1984). In the case of band-bending over a distance which is shorter than the De Broglie wavelength of about 100 Å for holes (in diamond), the three-dimensional (3D) density-of-states (DOS) changes to a two-dimensional (2D) DOS as shown schematically in Fig. 5b. For this calculation the hetero-junction effect is modeled using a graded interface in which the barrier height, as well as the effective mass is assumed to change smoothly in a transition layer whose thickness is specified. Holes move in an effective potential given by:

$$V(x) = e\phi(x) + V_h(x)$$

where $\phi(x)$ is the electrostatic potential, $V_h(x)$ is the effective potential energy associated with the hetero-junction discontinuity, which we assume to be 1.68 eV. The normalized envelope function $\zeta_i(x)$ for the hole sub-band i is given by the Schrödinger equation of the BenDaniel-Duke form:

$$-\frac{\hbar^2}{2}\frac{d}{dx}\frac{1}{m_n(z)}\frac{d\zeta_i(x)}{dx} + V(x)\zeta_i(x) = E_i\zeta_i(x) \qquad \text{(Eq. 7)}$$

where $m_n(x)$ is the position-dependent effective mass (n stands for: HH, LH, SO) and $E_i$ is the energy of the bottom of the i-th sub-band. The Poisson equation for the electrostatic potential takes the form:

$$\frac{d}{dz}\varepsilon_o\varepsilon_r(x)\frac{d\phi(x)}{dx} = e\sum N_i\zeta_i^2(x) - \rho_I(x) \qquad \text{(Eq. 8)}$$

with:

$$N_i = \frac{m_n kT}{\pi\hbar^2}\ln\left[1 + \exp\left(\frac{E_i - E_F}{kT}\right)\right] \qquad \text{(Eq. 9)}$$

where $\varepsilon_r(x)$ is the position-dependent dielectric constant, assumed to be constant in diamond ($\varepsilon_r$ = 5.7). For the adsorbate layer we varied $\varepsilon_r$ between 1 and 5.7. The calculations show that $\varepsilon_r$ does not affect the energy levels in the quantum well but does affect the width of the wave function of holes extending out of the diamond into the water layer. In the following we show results deduced for $\varepsilon_r$ = 5.7. $N_i$ is the number of holes per unit area in the sub-band i, $E_F$ is the Fermi-energy and $m_n$ represents the mass of holes (HH, LH, SO). As a first order approximation we neglect impurities ($\rho_I(x)$ = 0). Calculations have been performed for hole sheet-densities in the range 5x10$^{12}$ cm$^{-2}$ to 5x10$^{13}$ cm$^{-2}$ (Nebel, C.E.; Rezek, B.; Zrenner, A. 2004a and b).

A typical result is shown in Fig. 13. In diamond at the interface to the H-terminated surface, three discrete energy levels for holes govern the electronic properties, namely the first sub-bands of the LH-, HH- and SO-holes. For a hole sheet-density of $5 \times 10^{12}$ cm$^{-2}$, levels at 221 meV (HH), 228 meV (SO) and 231meV (LH) below the valence band maximum at the surface (VBM$_S$) are deduced. The Fermi-energy is 237 meV below VBM$_S$. Also shown are normalized hole wave-functions labeled LH, HH and SO. In thermodynamic equilibrium the chemical potential of the adsorbate layer and the Fermi-level of diamond are in equilibrium. Our calculations reveal an energy gap between electrons in the adsorbate layer and holes in the quantum well. In case of $5 \times 10^{12}$ cm$^{-2}$, holes must overcome 6 meV or more to recombine with electrons. The wave-function of holes extends about 5 Å into the adsorbate layer calculated for a dielectric constant $\varepsilon_r$ of 5.7. For a hole sheet-density of $5 \times 10^{13}$ cm$^{-2}$, three discrete energy levels are calculated, namely 770 meV (HH), 786 meV (SO) and 791 meV (LH) below the VBM$_S$. The first three sub-levels are occupied by holes and recombination is prevented by a gap of 90 meV.

Such two dimensional properties can be expected only theoretically as the real surface is governed by several additional properties like: 1) Surface and bulk defects which may pin the surface Fermi level. 2) Surface roughness. 3) Ions in the Helmholtz layer of the adsorbate film in close vicinity to the hole channel. These parameters may cause broadening of the 2D levels so that a continuous, semi-metallic DOS may be a better description (Nebel, C.E., Ertl, F., Sauerer, C., et al. 2002, Sauerer, C., Ertl, F., Nebel, C.E., 2001).

## 4.5 In-plane capacitance-voltage properties of Al on H-terminated diamond

The capacitance-voltage data shown in Figs. 11 and 12 indicate that a peripheral depletion layer is present between Al and H-terminated diamond that is covered with an adsorbate layer to generate transfer doping. This is schematically displayed in Fig. 9. Towards the Al contact, the accumulation layer is depleted as the work function of Al relative to H-terminated diamond is misaligned. This depletion region generates an in-plane capacitance which can be detected experimentally and which scales approximately with the length of the periphery (Fig. 12). As contacts were squares, some deviations from a perfect fit can be expected. The data indicate that in-plane Schottky-junction (2D) properties of Al on H-terminated diamond dominate the electronic properties.

General features of a junction between a two-dimensional electron gas and a metal contact with Schottky properties have been discussed by S.G. Petrosyan and A.Y. Shik 1989 and by B. Gelmont and M. Shur 1992. Following their arguments the capacitance of a metal in contact with a 2D gas is well described by:

$$C = \frac{\varepsilon_o \varepsilon_r L}{\pi} \ln \left\{ \frac{\left(d_{Al}^2 + x_{dep}^2\right)^{0.5} + d_{Al}}{\left(d_{Al}^2 + x_{dep}^2\right)^{0.5} - d_{Al}} \right\} \quad \text{(Eq. 10)}$$

where $L$ is the length of the metal periphery, $d_{Al}$ is the thickness of Al and $x_{dep}$ is the width of the space-charge region. Taking into account the thickness of Al as 600 nm, $\varepsilon_r$ = 5.7, and the detected variation of the capacitance, the variation of the depletion width in our experiments is in the range of 10 to 300 nm (see Fig. 14). The shortness of the depletion layer is a result of the high hole sheet-carrier density, which is in the range $10^{12}$ to $10^{14}$ cm$^{-2}$. S.G. Petrosyan and A.Y. Shik 1989 calculated an inverses proportional relationship between the width of the depletion layer and the sheet carrier density, given by:

$$x_{dep} \propto \frac{\varepsilon_o \varepsilon_r}{2\pi e} \frac{V}{p_{sh}} \qquad \text{(Eq. 11)}$$

where V is the applied voltage, e is the elementary charge, and $p_{sh}$ the sheet hole density. It is interesting to note that such an in-plane (2D) junction also follows the exponential law given by:

$$j \approx j_{rev} \left[ \exp(\frac{eU}{kT}) - 1 \right] \qquad \text{(Eq. 12)}$$

where $j_{rev}$ is the contact specific reverse current. To summarized these results briefly: The detected absolute values of the capacitance are orders of magnitude too small to be discussed using a parallel plate model. The most reasonable model is an in-plane Schottky model. Unfortunately, such junctions show comparable characteristics to three dimensional (conventional) Schottky junctions and are therefore not distinguishable if only IV-experiments are applied. Only a combination of CV and IV helps to elucidate the detailed properties.

### 4.6  Hole carrier propagation and scattering in the surface layer

To elucidate the transport properties of holes in the surface channel of diamond without scattering by grain boundaries and surface roughness, several undoped single crystalline CVD diamond films of 1µm thickness have been grown homoepitaxially on 3 mm x 3 mm (100) oriented synthetic Ib substrates, using microwave plasma chemical vapor deposition (CVD). Typical deposition parameters were sample temperature 800 °C, power 750 W, gas pressure 25 Torr, hydrogen flow 398 sccm and methane flow 2 sccm. To achieve H-termination after growth, the $CH_4$ is switch off and diamond is exposed to a pure hydrogen plasma for 5 minutes with otherwise identical parameter. Samples are then cooled down to room temperature in $H_2$ atmosphere. Temperature ramping, gas flow, and plasma switching are shown as function of time schematically in Fig. 15. A detailed discussion of growth and sample properties is given in a) Watanabe, H.; Takeuchi, D.; Yamanaka, S., 1999, b) Okushi, H., 2001, Okushi, H.; Watanabe, H.; Ri, R.S.; Yamanaka, S.; Takeuchi, D., 2002, and c) Watanabe, H., Ri, R.I.S. , Yamanaka, S., Takeuchi, D., Okushi, H. 2002. The high quality of these films is demonstrated by cathode-luminescence (CL) measurements and atomic force microscopy (AFM). In CL free-exciton emission is detected at 300 K at 5.27 eV and 5.12 eV, which is only the case for ultra-pure diamond (negligible extrinsic contaminations and intrinsic defects). In addition, AFM shows atomically flat surface properties so that the effect of surface roughness on carrier propagation should be negligible.

The samples were then characterized by Hall effect measurements (for details see Rezek, B,, Watanabe, H., and Nebel, C.E. 2006) which have been performed in ambient air at 300 K. Then the samples were annealed at 400 K at 1 Torr in argon gas for a given period of time. After that temperature dependent Hall experiments have been performed.

At room temperature, the mobility increased after 1 h annealing at 400 K from original 114 cm$^2$/Vs to about 223 cm$^2$/Vs. The sheet carrier concentration was strongly affected by thermal annealing: It decreased from 7x10$^{12}$ cm$^{-2}$ to 6x10$^{11}$ cm$^{-2}$. The sheet conductivity decreased from 1x10$^{-4}$ ($\Omega$/sq)$^{-1}$ to 2x10$^{-5}$ ($\Omega$/sq)$^{-1}$. Obviously, the increasing carrier mobility is accompanied by a decrease in carrier density and rise of resistivity.

Fig. 16 shows the detected variation of mobility and sheet hole density as a function of temperature. The mobility shows a maximum around 240 K, decreasing toward lower and higher temperatures. The sheet carrier concentration exhibits a monotonous increase toward high temperatures over the whole temperature range, with an activation energy of (9 – 23) meV. This is in good agreement with previously reported data which showed that transport properties of holes at hydrogenated diamond surfaces are near to metallic (Nebel, C.E., Sauerer, C., Ertl, F. et al. 2001, Nebel, C.E., Ertl, F., Sauerer, C. et al. 2002).

The time evolution of the mobility and sheet carrier concentration at the annealing temperature of 400 K is shown in Fig. 17. During annealing, the carrier concentration decays exponentially ~ exp(-$t/\tau$) (full line in Fig. 17 "sheethole density) with a time constant $\tau$ of (6.8±0.3) h. At the same time, the mobility increases, which is well fitted by an exponential increase of $\mu$ ~ (1-exp(-$t/\tau$) (dashed line in Fig. 17 "mobility") using the same time constant.

Figure 18 shows mobilities and carrier concentrations as measured at 300 K after annealing at 400 K for 1 to 16 h. Compared to the data at 400 K (see Fig. 17), a more pronounced increase of mobility (squares) with annealing time is detected with values reaching up to 328 cm$^2$/Vs after 16 h. The sheet carrier concentration (dots) decays in correlation with the increase in mobility.

Exposing the samples to air for several days results in a complete recovery of the original values of mobility, hole concentration, and conductivity. The recovery indicates that the annealing did not degrade the diamond surface H termination but rather modified the properties of the adsorbate layer on the diamond surface.

The most pronounced increase in mobility was enabled by mechanical cleaning step on hydrogenated surfaces. Prior to cleaning the mobility of holes at room temperature was 92 cm$^2$/Vs in air and increased to 115 cm$^2$/Vs after 1 h annealing in vacuum. Using mechanical cleaning, the mobility was typically 114 cm$^2$/Vs in air and increased further to 223 cm$^2$/Vs after 1 h annealing in vacuum. The sheet carrier concentration decreased after cleaning by about a factor of 1.5.

As illustrated in Fig. 1, the mobility decreases toward lower or higher temperatures from a maximum centered at about 240 K. The decrease is remarkably close to the tendencies corresponding to ionized impurity scattering ~ T$^{+3/2}$ toward lower temperatures and phonon scattering ~ T$^{-3/2}$ toward higher temperatures (Seeger, K. 1999). These power-law tendencies were introduced for bulk three-dimensional scattering processes. Yet, even in 2D systems the carrier transport is dominated by three-dimensional effects if carrier mobilities are relatively low and typically < 1000 cm$^2$/V s (Ando, T., Fowler, A.B., and Stern, F. 1982).

Considering the transfer doping model of hydrogenated diamond the exponential decay of carrier concentration with annealing time indicates thermal desorption of surface adsorbates. Since the time constant of the desorption process is expressed as $\tau = (1/\nu_o)\exp(E_d/kT)$ the desorption energy can be calculated by:

$E_d = kT \ln(\tau \nu_o)$   Eq. 13

Using τ = 6.8 h and typical attempt-to-escape frequency prefactor $\nu_o = 10^{13}$ s$^{-1}$, Eq. 1 yields $E_d$=1.1 eV. This is about 2 times larger than typical desorption energies which are in the range of 0.5 eV, typical for physisorbed water molecules (Chakarov, V., Österlund, L, and Kasemo, B. 1995). The additional energy may arise from the fact that the decrease of hole concentration in the surface channel requires not only desorption but also transfer of electrons from adsorbate layer back into diamond via an electrochemical reaction (Maier, F., Riedel, M., Mantel, B. et al 2000).

Based on our data, we assume that due to the small separation of few Angstroem between holes in the surface conductive channel and ions in the adsorbate layer on the surface (Nebel, C.E., Rezek, B., and Zrenner, A. 2004a and b) holes are scattered by electrostatic interaction with ions. Thus, the annealing leads to higher mobilities by reducing the concentration of ionized adsorbates at the surface.

# 5. Surface electronic properties of diamond in electrolyte solutions

## 5.1 Redox couple interactions with undoped H-terminated CVD diamond

Electrochemical experiments on H-terminated diamond are suited to investigate transfer doping properties as the electrolyte solutions are well defined with respect to pH and redox couple chemical potentials (Nebel, C.E., Rezek, B., Shin, D., Watanabe, H., Yamamoto T., 2006, Nebel, C.E., Kato, H., Rezek, B. et al 2006, Shin, D., Watanabe, H., Nebel, C.E., 2005, Shin, D., Watanabe, H., Nebel. C.E. 2006). To characterize these properties the same undoped hydrogen terminated CVD diamond substrate has been applied as electrode as discussed above in Hall effect experiments. The contact arrangement, shown in Fig. 19, has been optimized for the electrochemical experiments. 200 nm thick Au films have been evaporated on the H-terminated surface with rotational symmetry. The electrochemical active area is about 0.8 mm$^2$. Pt is used as counter electrode, a saturated calomel electrode (Hg/Hg$_2$Cl$_2$) as reference electrode and a voltammetric analyzer is applied for cyclic voltammetry experiments. The scan rate is 100 mV/s. To calculate the RC-time limitation of the experiments the active area of 0.8 mm$^2$, the experimentally detected series resistance of the surface conductive film of $10^5$ Ω, and the typical double layer capacitance of 5 μF/cm$^2$ as also detected by M.C. Granger et. al. (Granger, M.C., Xu, J., Strojek, J.W., Swain, G.M., 1999) has been taken into account. Using τ = RC, results in a time constant of about 4 ms. This is two to three orders of magnitude faster than our potential scan rate of 100 mV/s. We exclude therefore RC-limited effects in our cyclic voltammetry experiments.

Redox analyte of 10 mM concentration has been mixed to the supporting 0.1 M Na$_2$SO$_4$ electrolyte solution. As redox analytes we used well characterized molecules like Fe(CN)$_6^{3-/4-}$ (formal potential ($U_{Fe}^o$) is +0.46 V vs. SHE, chemical potential μ$_{Fe}$ = -4.9 eV below the vacuum level), Ru(NH$_3$)$_6^{2+/3+}$ ($U_{Ru}^o$ = +0.025 V vs. SHE, μ$_{Ru}$ = -4.46 eV), methyl viologen MV$^{2+/1+}$ ($U_{MV2}^o$ = -0.48 V vs. SHE, μ$_{MV2}$ = -3.96 eV), MV$^{+1/0}$ ($U_{MV1}^o$ = -0.81 V vs. SHE, μ$_{MV2}$ = -3.63 eV) and Co(sep)$^{2+/3+}$ ($U_{Co}$ = -0.38 V vs. SHE, μ$_{Co}$ = -4.06 eV) (Granger, M.C., Witek, M., Xu, J. et al. 2000). The redox energy levels with respect to vacuum and to H-terminated diamond are shown in Fig. 20.

Cyclic voltammetry can be used to gather information about diamond film and surface qualities. In particular, this method can be applied to detect non-diamond carbon which is present at grain boundaries or graphitic deposits at the surface. It is known that sp2 gives rise to catalytic effects on hydrogen and oxygen evolution which reduces the working potential window significantly. Fig. 21 shows background cyclic voltammetry IV-curves as detected in 0.1 M $H_2SO_4$ (pH = 1) on our sample (DRC, full line in Fig. 21). The data are compared with two high quality polycrystalline boron doped diamond films from the literature (Granger, M.C., Xu, J., Strojek, J.W., Swain, G.M., 1999, Granger, M.C., Witek, M., Xu, J. et al. 2000).

On our undoped, H-terminated CVD diamond the electron exchange reaction between the hole accumulation layer and the electrolyte reveals distinct features. A large chemical potential window extends from less than -3 V (detection limit of our set-up) to +1.6 V (for pH 1). In this regime the background current density is in the $\mu A/cm^2$ range, as can be seen in the inset of Fig. 21 where the background current is plotted in log scale. The general feature is comparable to a diode. We attribute the large window and the small current density firstly to the absence of $sp^2$ carbon as no grain boundaries are at the surface (Angus, J.C., Pleskov, Y.V., Eaton, S.C., 2004, Granger, M.C., Witek, M., Xu, J. et al. 2000, Swain, G.M., 2004, Fujishima, A., Einaga, Y., Rao, T.N., Tryk, D.A. 2005). Hydrogen evolution cannot be detected down to -3V, which is the limit of our set-up. Secondly, we attribute it to the fact that for negative potentials, applied to the diamond film, the surface conductivity disappears (see next chapter) as the hole-accumulation layer becomes depleted.

For positive potentials the surface conductivity remains. Therefore, the diamond surface can be oxidized at potentials larger than +1.6 V at pH 1, which removes hydrogen from the surface but does not produce any morphological damage to the surface.

A summary and comparison of redox reactions with H-terminated diamond is shown in Fig. 21. For $Fe(CN)_6^{3-/4-}$ and $Ru(NH_3)_6^{2+/3+}$ electron transfer from the redox couple into the diamond electrode (oxidation peak) and from the diamond to the redox couple (reduction peak) can be clearly detected. For methyl viologen $MV^{2+/1+}/MV^{+1/0}$ and $Co(sep)^{2+/3+}$ the result indicates no redox interaction.

To discuss these in a more general context, we compare our data with data published in the literature for H-terminated highly boron doped ($10^{19} – 10^{20}$ $cm^{-3}$) diamond (Granger, M.C., Xu, J., Strojek, J.W., Swain, G.M., 1999). These films are three dimensional electrodes in contrary to our in-plane contact geometry. Cyclic voltammetry experiments on these polycrystalline films have been performed using 1M KCl electrolyte solution and 0.05 V/s. To overcome problems which arise by comparison in absolute terms, we discuss in the following normalized data as shown in Fig. 23 (a: $Fe(CN)_6^{3-/4-}$, b: $Ru(NH_3)_6^{2+/3+}$, c: $Co(sep)^{3+/2+}$), d: $MV^{2+/1+}/MV^{+1/0}$).

For $Fe(CN)_6^{3-/4-}$ and for $Ru(NH_3)_6^{2+/3+}$, hydrogen terminated intrinsic diamond acts like a metal electrode, with oxidation peak amplitudes of about 0.37 $mA/cm^{-2}$ ($Fe(CN)_6^{3-/4-}$) and 0.46 $mA/cm^2$ ($Ru(NH_3)_6^{2+/3+}$). $\Delta U_{pp}$ for $Fe(CN)_6^{3-/4-}$ is 296 mV and for $Ru(NH_3)_6^{2+/3+}$ 365 mV compared to 64 mV and 65 mV as detected for metallic polycrystalline diamond (Granger, M.C., Xu, J., Strojek, J.W., Swain, G.M., 1999). The oxidation/reduction currents are reversible. The peak-shift and broadening indicate rate limited electron transfer as we exclude RC-time constant effects. This is well described in the literature based on the Marcus-Gerischer model (Gerischer, H., 1961). For $Fe(CN)_6^{3-/4-}$ and Ru(NH$_3$)$_6^{2+/3+}$ the reduction and oxidizing currents are symmetrically centered around the formal potential U° of +224 mV (Fe(CN)$_6^{3-/4-}$) and -230 mV (Ru(NH$_3$)$_6^{2+/3+}$). This is in good agreement with the data of Granger et al. (Granger, M.C., Xu, J., Strojek, J.W., Swain, G.M., 1999) who detect +220 mV for Fe(CN)$_6^{3-/4-}$ and -217 mV for Ru(NH$_3$)$_6^{2+/3+}$. Taking into account the relation between electrode potential and electron energy with respect to the vacuum level we use (Angus, J.C., Pleskov, Y.V., Eaton S.C., 2004):

$\varepsilon = (-e)U° - 4.44$ eV        (Eq. 14)

where $\varepsilon$ is the energy of electrons with respect to the vacuum level, e is the elementary charge, U° is the formal potential (vs. SHE), and 4.44eV is the scaling parameter between electrode potential and electron energy. The chemical potential of redox couples Fe(CN)$_6^{3-/4-}$ and Ru(NH$_3$)$_6^{2+/3+}$ are calculated to be -4.9 eV (Fe(CN)$_6^{3-/4-}$) and -4.45 eV (Ru(NH$_3$)$_6^{2+/3+}$) which results in Fermi level positions of about 530 meV and 80 meV below the valence-band maximum at the surface, assuming a negative electron affinity of about -1.1 eV (see Fig. 20). For these two redox couples transfer doping conditions are fulfilled which gives rise to the insulator-metal transition of intrinsic diamond.

Cyclic voltammetry experiments on methyl viologen (MV$^{2+/1+}$/MV$^{+1/0}$) and Co(sep)$^{3+/2+}$ show distinct differences. The redox acitivities are shifted deep into negative potentials (with respect to diamond electrode). Here the surface conductive layer is depleted and diamond becomes an insulator.

Application of potentials larger than +1.6 V gives rise to oxidation of hydrogen terminated diamond surfaces. To characterize to electronic properties in the transition region from perfectly hydrogen terminated to fully oxidized, we have performed a sequence of voltammetric experiments using 10 mM Fe(CN)$_6^{3-/4-}$ as redox couple in 0.1 M Na$_2$SO$_4$ electrolyte solution (pH = 6). The scan rate of the voltammetric experiment was 0.1 V/s in the regime -1.2 V to +1.6 V. The results are shown in Fig. 24. The application of a sequence of short time anodic oxidation can be summarized as follows:

1) The reduction peak marked by "a" in Fig. 24 is decreasing rapidly, while the oxidizing peak remains nearly stable (steps 1 to 4).

2) After the disappearance of the reduction peak, the oxidizing peak, marked "b" in Fig. 24, is decreasing and shifting to higher potentials (steps 5 to 12).

3) Finally, the redox-interaction completely disappears but oxygen evolution still can be detected. We attribute this to the chemical reaction:

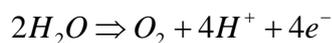

$$2H_2O \Rightarrow O_2 + 4H^+ + 4e^-$$

which has an energy level at -5.3 eV below the vacuum (Angus, J.C., Pleskov, Y.V., Eaton S.C., 2004). This is significantly deeper than for Fe(CN)$_6^{3-/4-}$ (energy level: -4.9 eV).

4) After prolonged anodic oxidation the current approaches zero due to the complete removal of hydrogen and of the surface conductivity.

## 5.2 Electrochemical exchange reactions of H-terminated diamond with electrolytes and redox couples

CVD diamond films under discussion are perfect insulators if measured in vacuum with a clean surface. In this case the Fermi-level is in the band gap of diamond. The same films become metallic if immersed into redox-electrolyte solutions with chemical potentials below the valence band maximum. This can be ascribed to alignment of Fermi-level and redox-potential or chemical potential at the interface. It is a remarkable phenomenon, which requires "defect free" bulk properties and perfect hydrogen termination of the surface carbon dangling bonds to unpin the Fermi-level. Both conditions seem to be fulfilled for these samples. Obviously, optimized growth and hydrogen termination of diamond generates high quality sub-surface and surface electronic properties. It is also certain that the insulator metal transition is related to the presence of hydrogen at the surface as the electrochemical currents vanish after anodic oxidation.

A comparison with metallic boron doped polycrystalline diamond shows that the current voltage variations are broader and that the maxima are shifted to higher (oxidation) or lower potentials (reduction) with respect to the formal potential. Based on arguments given above we exclude RC-limitations as origin for the shifts and broadening. We assume that these variations are generated by limited electron transfer rates at the interface. This is schematically illustrated in Fig. 25. Metallic properties are established by occupied and empty electronic states in the valence-band, separated by $E_F$. In case of metals, the density-of-state distribution $\rho(E)$ is approximately constant around $E_F$ whereas for diamond $\rho(E)$ is a function of energy. The application of external potentials shifts $E_F$, either up (negative potential) by filling of empty electronic levels or down (positive potential) by emptying of valence-band states. External potentials will, however, not change the confinement of holes (band-bending) significantly, as potentials are applied parallel to the conductive layer (perpendicular to the band bending).

The interaction of metal and semiconducting electrodes with redox couples has been described in the literature using the Marcus-Gerischer model (Bard, A.J., Faulkner, L.R. 2002), where the oxidizing density-of-states ($D_O(E)$) above the chemical potential $\mu$ and the reduction states ($D_R(E)$) below $\mu$ are Gaussian shaped. An upward shift of the Fermi energy, $E_F$, gives rise to electron transfer from the electrode into empty Do states and vice versa. The local rate of reduction $r_R(E)$ or oxidation $r_O(E)$ are given by:

$$r_R(E) = A\{f(E)\rho(E)\varepsilon_{red}(E)D_O(E)dE\}/\Delta t \quad \text{(Eq. 15)}$$

$$r_O(E) = A\{[1-f(E)]\rho(E)\varepsilon_{ox}(E)D_R(E)dE\}/\Delta t \quad \text{(Eq. 16)}$$

where A is the electrode area, $\Delta t$ a given time interval, $f(E)$ the Fermi-distribution, $\rho(E)$ the density-of-state distribution (DOS) of the metal or semiconductor, $D_O(E)$ the DOS of the oxidizing redox states, $D_R(E)$ the DOS of the reduction states, $\varepsilon_{red}(E)$ and $\varepsilon_{ox}(E)$ the proportionality functions for reduction (red) and oxidation (ox), and dE the energy interval under discussion. The exchange rates are proportional to the DOS of the electrode $\rho(E)$.

The transition rate equations contain $\varepsilon_{red}(E)$ and $\varepsilon_{ox}(E)$. These parameters are governed by the wave-function overlap of electrons in the valence-band of diamond with empty $D_O$ states and of electrons in the $D_R$ states with empty states in the valence-band of diamond. Here, the hydrophobic properties of H-terminated diamond may play an important role as the gap between the redox-electrolyte molecules and the diamond surface will be a function of wetting characteristics. Holes in the surface conducting channel are confined by a dipole energy barrier of about 1.7 eV (Ristein, J., Maier, F., Riedel, M., Cui, J.B., Ley, L., 2000). This conductive layer is very thin (10-20 Å), the continuous density-of-state distribution may therefore be transformed into a two-dimensional (2D) system with discrete energy levels. It is however very likely, that the 2D-states are broadened by disorder due to ionic interface scattering and surface roughness.

The experiments show that for two redox couples with chemical potential below the valence-band maximum oxidation and reduction currents can be detected with formal potentials like for H-terminated metallic polycrystalline boron doped diamond. This indicates that the electrochemical interaction of boron doped diamond is also governed by energy alignment properties which are dominated by hydrogen termination and related negative electron affinities. The conductivity of a 3D electrode (metallic boron doped diamond) is, however significantly higher than that of a conductive surface layer of 10 to 20 Å thickness. It affects the dynamic properties of cyclic voltammetry experiments as shown in Fig. 23 and 24. In case of a 3D metallic boron doped film the peaks are narrow and separated only by about 65 mV. We attribute the different results detected for methyl viologen and Co(sep)$^{3+/2+}$ to the depletion of the surface hole accumulation layer at potentials where the redox activity is expected.

Finally we want to address the oxidation phenomena as detected by cyclic voltammetry. It is known that H-terminated diamond has a negative electron affinity in the range -1.1 to -1.3 eV and oxidized diamond shows a positive electron affinity. We expect therefore a vacuum level shift at the surface of diamond of more than $\Delta E \geq 1.1$ eV. Our data indicate that partial oxidation generates a partial shift of the vacuum level, which gives rise to a redox-level shift as shown schematically in Fig. 26. After first partial oxidation the "reduction peak" (electron flow from the diamond to the redox-couple) disappears. This indicates that electrons in the surface conducting layer of diamond cannot be transferred into the redox couple as only occupied $D_R(E)$ states are aligned with the conductive layer. Further oxidation shifts the redox levels even higher, so that the "oxidizing peak" also decreases and finally vanishes. The redox-couple and the electronic levels of the valence band are approaching a situation which normally prevents the formation of a conductive layer as the redox energetic layers are higher than the valence band maximum of diamond. As the redox couple is dissolved in an electrolyte of 0.1 M of $Na_2SO_4$ with pH 6, the chemical potential of the oxygen evolution at -5.3 eV, given by the reaction:

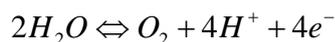

$$2H_2O \Leftrightarrow O_2 + 4H^+ + 4e^-$$

can accommodate electrons ("formation of surface conductivity") if $O_2$ is in the electrolyte. This is the case in our experiments as we applied anodic oxidation potentials (> +1.6 V). Fig. 26 shows the energy diagram schematically. As long as this level is below the valence band maximum $E_{VBM}$, surface conductivity can be expected. Prolonged oxidation gives rise to insulating surface properties, as all H is chemically removed from the surface. As a result the positive electron affinity misaligns energy levels significantly which prevents transfer doping and redox activity. It is interesting to note that a similar result has been reported by Cui and co-workers (Cui, J.B., Ristein, J., Ley, L., 1998). They applied photoelectron emission experiments on hydrogen terminated diamond and removed the hydrogen termination partially by a sequence of annealing steps. They also find a continuous shift of the vacuum level from -1.3 eV to +0.38 eV for a hydrogen free surface. This is in good agreement with our electrochemical results where we removed hydrogen by a sequence of oxidizing steps.

### 5.3 Ion sensitive field effect transistor (ISFET) from undoped CVD diamond

In the following we address the pH sensitivity of the surface conducting layer as last part of the transfer doping phenomena. To measure the variation of surface conductivity in different pH-liquids, we manufactured ion sensitive field effect (ISFET) structures (Kawarada, H., Araki, Y., Sakai, T. et al. 2001, Rezek, B., Watanabe, H., Shin, D., Yamamoto, T., Nebel, C.E., 2006). The geometric properties are shown in Fig. 27 where a hydrogen terminated surface area of 100 $\mu$m x 500 $\mu$m size connects two Au contacts of 130 $\mu$m x 130 $\mu$m which have been thermally evaporated on to diamond. Au wires are bonded to these pads and insulated by a resistive lacquer from contact to electrolyte solution.

The outer part has been oxidized by a soft oxygen plasma to generate insulating surface properties. The schematic arrangement of ISFET in electrolyte solution is shown in Fig. 28. Platinum is used as "semi-reference" electrode. The level of pH was varied between 2 to 12 by use of Britton-Robinson buffer (0.04M $H_3BO_3$+$H_3PO_4$+$CH_3COOH$, titrated by NaOH (0.2M)).

Variations of the drain-source currents with pH, measured at constant drain-source potentials of 0.2 V are shown in Fig. 29. The currents show transistor characteristics with increasing amplitude for decreasing pH. For positive gate potentials the current is pinched off with thresholds, dependent on pH. This supports the results of Chaper III.2 where the redox activity of Co(sep) and Methyl-Viologen has been quenched. To prevent degradation by partial oxidation, the maximum applied negative gate voltage was limited to -0.5 V. At this gate potential the drain-source resistance is in the range (7 - 13)x$10^4$ $\Omega$. Defining 1 $\mu$A drain source current as reference, a pH sensitivity of 66 mV/pH is calculated as shown in the inset of Fig. 29. This is slightly higher than expected by the Nernst law (59 mV/pH). From the leakage current of the ISFET (current flow from surface channel of diamond into the electrolyte) which is about 3 $\mu A/cm^2$ at -3 V, we calculate a gate resistance of $10^8$ $\Omega$. The drain source resistance is in the range of $10^5$ $\Omega$ (Nebel, C.E., Rezek, B., Shin, D., Watanabe, H., Yamamoto, T., 2006).

The gate properties can be best described by diode characteristics. Below a critical gate/channel voltage there is only a small leakage current flowing. Above the threshold voltage the diode opens and exponentially increasing currents can be detected. The electronic circuit of a diamond ISFET is summarized in Fig. 30. Gate leakage threshold voltages are shown in Fig. 31 for different pH, measured in 0.1 M $HClO_4$ (A) (pH 1), 0.1 M $Na_2SO_4$ (B) (pH 6), and 0.1 NaOH (C) (pH 13). The threshold voltage of oxygen evolution at positive potentials ($U_{OX}$) shifts from +0.7 V (vs. SCE) in pH 13 to about +1.6 V in pH 1. The shift is about 60 mV/pH, which is in reasonable agreement with the Nernst prediction. Oxidation of H-termination occurs for potentials lager than $U_{OX}$ which limits the application of pH sensors to gate potentials smaller than 0.7 V.

# 6. Conclusions

The combination of results from ISFET characterization and from cyclic voltammetry experiments on the very same samples show, that a surface conducting layer is generated in all investigated pH-electrolytes, ranging from 2 to 12. The drain-source currents of H-terminated single crystalline diamond ISFETs show high pH sensitivity, with about 66 mV/pH. The basic mechanism behind this phenomenon is the alignment of chemical potential and Fermi level of diamond ("transfer doping model") as shown schematically in Fig. 32a. In case of a low pH, the chemical potential with respect to the vacuum is large and therefore the Fermi level is deep in the valence band, giving rise to a high drain source current (Fig. 32a), which is in agreement with our findings. With increasing pH the chemical potentials shifts up in energy as well as $E_F$ and the drain source current is decreasing. For the site-binding model, where a partially oxidized surface is required (Kawarada, H., Araki, Y.,Sakai,T., Ogawa, T., Umezawa, H., 2001, Madou, M.J., Morrison, S.R., 1989) opposite characteristics are detected (see Fig. 32b). We therefore conclude that our pH sensitivity favors the transfer doping model.

The formation of surface conductivity in diamond requires electron transitions from occupied valence-band states into empty states of the electrolyte. This is possible if we follow the arguments of Angus et al. (Angus, J.C., Pleskov, Y.V., Eaton, S.C., 2004) and Ristein et al. (Ristein, J., Riedel, M., Ley, L., 2004) who assume that electrons will trigger the reduction of atmospheric oxygen or other atmospheric contaminants in the electrolyte, described by: $O_2 + 2H_2O + 4e^- \rightarrow 4OH^-$. If we discuss this phenomenon like a redox couple interaction ($O_2/OH^-$), the $O_2$ is the empty state of the couple and energetically slightly above $OH^-$. As the energy level of $OH^-$, and therefore of $O_2$, for all pH is below the valence-band maximum the formation of a surface conducting film will take place in all electrolytes. The reaction will be limited either by a complete consumption of $O_2$ and/or by the electric field which builds up at the surface and which prevents unlimited electron flow. The consumption argument is of minor importance if we take into account absolute numbers. Considering data from Hall effect experiments we know that the typical sheet hole density is in the range of $10^{12}$ cm$^{-2}$ for drain-source conductivities like in our experiments. The ISFET sensor area is $4\times10^{-4}$ cm$^2$, which results in about $10^8$ electrons which are exchanged at the interface to reduce $O_2$. This is a small number which may not generate saturation effects.

In summary, the formation of a conductive layer at the surface of undoped diamond has been detected and characterized by Hall effect, conductivity, contact potential difference (CPM), scanning electron microscopy (SEM), and cyclic voltammetry experiments. The revealed data show that due to electron transfer from the valence band into empty states of an electrolyte or adsorbate, a highly conductive surface layer is generated. Holes propagate in the layer with mobilities up to 350 cm$^2$/Vs. The sheet hole density in this layer is in the range $10^{11}$ to $5\times10^{12}$ cm$^{-2}$, and dependents on pH of the electrolyte or adsorbate. This can been utilized to manufacture ion-sensitive field effect transistors (ISFET), where the drain source conductivity is pH dependent, with about 66 mV/pH. Application of potentials larger than the oxidation threshold of +0.7 V (pH 13) to +1.6 V (pH 1) gives rise to exponentially increasing currents between diamond and electrolyte and to partial surface oxidation.

In general, the electronic interaction of diamond with redox couples and pH electrolyte solutions is well described by the transfer doping model.

Theoretical calculations of the "ideal surface electronic properties" of H-terminated diamond in contact with an electrolyte reveal 2D properties, with discrete energy levels. The calculated energies are in reasonable agreement with contact potential difference experiments applied. Extended energy bands with 2D properties could be expected for atomically smooth diamond surfaces un ultra-pure CVD diamond. However, ions in the Helmholtz layer of the adsorbate film will cause strong Coulomb scattering as they are only a few Angström away from the hole-accumulation layer.

## Acknowledgements

The author likes to thank Dr. B. Rezek, Dr. D. Shin and Dr. H. Watanabe for their scientific contributions. He also wants to acknowledge the help of T. Yamamoto for fabricating ISFET structures and greatly appreciated discussions with Dr. Okushi, Dr. Yamasaki and with all members of the Diamond Research Center at AIST, Japan. Special thank goes to Dr. Park who supported these activities with the donation of an electrochemical analyzer. The authors also thank Elsevier, New Diamond and Frontier Carbon Technology, Whiley and American Physical Society to grant copyright permission for several figures published in related journals.

# Figure Captions

Fig. 1:

Top: Band scheme and electron affinity $\chi$ for the bare, the hydrogenated, and the oxidized diamond (100) surface. Bottom: Sketch of the atomic arrangement of the bare, the hydrogenated and the oxidized (100) diamond surface.

Fig. 2:

The electron affinity $\chi$ (solud squares)as a function of annealing time at 1000 K. The dashed line through the data points for $\chi$ is the result of a fit assuming first order desorption kinetics of hydrogen. The full line marks the transition from NEA to PEA.

Fig. 3:

Electron affinity $\chi$ versus oxygen coverage; the solid line is a fit to the data with a dipole moment $p_z$ = -0.10 e Å and a polarizability $\alpha$ = 2.0x10$^{-30}$Asm$^2$/V (for details see: Maier, F., Ristein, J. and Ley 2001).

Fig. 4:

Summary of hole mobilities and sheet-densities as detected at room temperature on a variety of films (dashed line is "guiding the eyes") (Data from Nebel, C.E., Rezek, B., Shin, D. 2006).

Fig. 5:

Schematic description of the diamond/adsorbate hetero-junction (a) non-equilibrated and (b) equilibrated. Electrons from the valence-band tunnel into empty electronic states of the adsorbate layer as long as the chemical potential $\mu_e$ is lower than the Fermi energy $E_F$.

Fig. 6:

Chemical potentials referred to the standard hydrogen electrode and semiconductor energies referred to the vacuum level. The electron energies for the couple $O_2 +4H^+ +4e^- = H_2O$ at pH 0 and 14 are shown along with the band edges for hydrogen terminated and oxidized diamond. The negative electron affinity has been assumed to be -1.1 eV on H-terminated diamond (Angus, J.C., Pleskov, Y.V., Eaton, S.C. 2004).

Fig. 7a:

Scanning electron microscopy image of diamond which has been partially covered with Au (A), and which is hydrogen terminated (B) and oxidized (C).

Fig. 7b:

Two-dimensional contact potential measurement (CPD) on the same sample as shown in Fig. 7a.

Fig. 7c:

The white line in Fig. 7b indicates the scan position of the spatial CPD profile shown here. Note, within experimental accuracy, no contact potential difference between Au and H-terminated diamond can be detected.

Fig. 8:

Schematic energy band diagram of the interface of H-terminated diamond covered with Au or Al. The $\Delta E_{Al}$ and $\Delta E_{Au}$ refer to the valence band maximum at the surface, $E_{VBM-SURF}$.

Fig. 9

Schematic surface energy diagrams of H-terminated diamond covered with an adsorbate layer which is in contact with aluminum and with gold. The data was calculated using contact potential difference experiments and assuming a negative electron affinity of -1.1 eV.

Fig. 10:

Two IV characteristics measured on Al/H-terminated diamond Schottky-junctions in air at T = 300 K with 250 μm x 250 μm contact size.

Fig. 11:

Capacitance-voltage data detected on three contact configurations with Al areas 250 x 250 μm$^2$, 100 x 100 μm$^2$ and 50 x 50 μm$^2$.

Fig. 12:

Capacitance of the Al contacts with sizes 50 x 50 μm$^2$, 100 x 100 μm$^2$ and 250 x 250 μm$^2$ plotted for U = 0, 1, 2 and 3 V as a function of periphery length (dashed line is guiding the eyes of a linear dependence).

Fig. 13:

Energy band diagrams at the interface of hydrogen terminated diamond and an adsorbate layer calculated for a hole sheet-density of $5\times10^{12}$ cm$^{-2}$. The energies refer to the valence band maximum at the interface (VBM$_{INT}$). The figure shows the calculated energy levels of the first sub-bands of the light hole (LH), heavy hole (HH) and split-off band (SO). Also shown are the normalized wave functions $\zeta$ of holes (LH, SO, HH).

Fig. 14:

Calculated capacitance variations as a function of width of the depletion layer. Experimentally the capacitance is in the range 0.06 pF to 1 pF, which shows that the depletion layer width varies between 10 and 300 nm.

Fig. 15:

Temperature and gas flow variations, plasma switching cycles for growth or H-termination of CVD diamond.

Fig. 16:

Temperature dependent Hall mobility (squares) and sheet hole concentrations (dots) as measured on diamond surface in 1 Torr vacuum after 1 h annealing at 400 K. Original values measured in air at 300 K are also shown.

Fig. 17:

Hall mobility and sheet hole density as measured in a vacuum at the annealing temperature 400 K as a function of time. Curves represent a fit of the exponential decay of carrier concentration (full line in sheet hole density) and an exponential increase of carrier mobility (dashed line in mobility).

Fig. 18:

Hall mobility and sheet hole density as measured in vacuum (1Torr) at 300 K as a function of annealing time. Curves are for guiding the eyes.

Fig. 19:

Contact configuration applied to characterize undoped H-terminated diamond by electrochemical experiments. The geometry is in-plane where Au is evaporated on H-terminated diamond to generate an Ohmic contact on a 1 μm intrinsic CVD diamond grown on Ib diamond substrate.

Fig. 20:

Redox chemical potentials of different redox couples with respect to the vacuum level. We assumed a negative electron affinity of -1.1 eV. Only $Fe(CN)_6^{3-/4-}$ and $Ru(NH_3)_6^{2+/3+}$ have levels below $E_{CBM}$.

Fig. 21:

Background cyclic voltammetry IV-curves as detected in 0.1 M $H_2SO_4$ (pH = 1) on a hydrogen terminated undoped single crystalline CVD diamond film (DRC, full line). The data are compared with two high quality polycrystalline boron doped diamond films (NRL and USC) from the literature (Granger, MC., Xu, J., Strojek J.W. et al. 1999). The inset shows the DRC-data in a log scale. The chemical window if DRC is larger than 4.6 V.

Fig. 22:

Cyclic voltammetric I-V curves for $Fe(CN)_6^{3-/4-}$, $Ru(NH_3)_6^{2+/3+}$, $Co(sep)^{3+/2+}$ and methyl viologen $MV^{2+/1+}/MV^{+1/0}$. After anodic oxidation the current decreased to zero (dashed line). Pt was used as counter electrode, a saturated calomel electrode ($Hg/Hg_2Cl_2$) as reference electrode and cyclic voltammetry experiments have been performed with scan rates of 0.1 V/s. Redox analyte of 10 mM concentration has been mixed to the supporting 0.1 M $Na_2SO_4$ electrolyte solution.

Fig. 23:

Normalized I-V curves for $Fe(CN)_6^{3-/4-}$ (a), $Ru(NH_3)_6^{2+/3+}$ (b), $Co(sep)^{3+/2+}$ (c), and methyl viologen (d) as detected on intrinsic H-terminated diamond (full squares) and on boron-doped (metallic) H-terminated polycrystalline CVD diamond (open circles) (Granger, MC., Xu, J., Strojek J.W. et al. 1999, Granger, M.C., Witek, M., Xu, J. 2000). In d), the open circles show cyclic voltammetric IV curves as measured on boron doped diamond using methyl viologen. The cycles are slightly changing and are discussed in detail in Ref. Granger, MC., Xu, J., Strojek J.W. et al. 1999.

Fig. 24:

Electronic properties in the transition region from perfectly hydrogen terminated to fully oxidized. A sequence of voltammetric experiments using 10 mM $Fe(CN)_6^{3-/4-}$ as redox couple in 0.1 M $Na_2SO_4$ electrolyte solution (pH = 6) has been applied (indicated by:1 to 13).

Fig. 25:

Schematic comaprison of a metal and a diamond electrode in contact with a redox system with chemical potential $\mu_e$ below $E_{VBM}$. We show the situation for a negative voltage applied to diamond which shifts the Fermi level $E_F$ up compared to $\mu_e$. This gives rise to a reduction electron flow into empty states of the redox couple.

Fig. 26:

Diamond and redox-couple interaction ($D_O(E)$ and $D_R(E)$) as a function of a) hydrogen termination, b) partially oxidized and c) fully oxidized. $D_e(E)$ is the chemical potential of the reaction $2H_2O <=> O_2 + 4H^+ + 4e^-$ the 0.1 M $Na_2SO_4$ electrolyte solution (pH = 6) which has an energy level of -5.3 eV below the vacuum.

Fig. 27:

Top view of a realized ISFET structure as measured by electron microscopy. The length of the sensitive H-terminated area is 500 µm and the width 100 µm. On both ends, Au contacts have been thermally evaporated which act as drain and source electrodes, contacted by wire bonding. The H-terminated area is electrically insulated from the surrounding by an oxidized diamond surface. The Au contacts and wires are covered by a lacquer, which cannot be seen in this figure.

Fig. 28:

Schematic arrangement of an ion-sensitive field effect transistor from diamond. Due to hydrogen termination of diamond a thin surface conductive layer is formed at the interface to the electrolyte. Au contacts serve as drain and source electrodes. Both are covered for insulation in the electrolyte solution by a lacquer.

Fig. 29:

Drain source currents as a function of gate voltage of an ISFET structure on H-terminated diamond, measured in electrolyte solutions with pH in the regime 2 to 12. The pH sensitivity is shown in the inset for $I_{DS}$ = 1 µA, which results in a pH sensitivity of 66 mV/pH.

Fig. 30:

The diamond/electrolyte interface described in electronic terms. The interface is dominated by the Helmholtz capacitance, a leakage resistance in the range of $10^8$ Ω, and diode properties if a critical threshold voltage for anodic oxidation is applied. The drain source resistance is in the range $10^5$ Ω.

Fig. 31:

Cyclic voltammograms as detected on a H-terminated undoped diamond electrode in (A) 0.1 M HClO$_4$, (B) 0.1 M Na$_2$SO$_4$, and (C) 0.1 NaOH. The oxidizing threshold shifts with increasing pH to lower potentials with about 60 mV/pH.

Fig. 32:

Schematic comparison of transfer doping (a) and site-binding (b) model. In case of a) the shift of the chemical potential deeper in energy with decreasing pH gives rise to an increase in surface conductivity, because the Fermi level also moves deeper into the valence band. For the site-binding model (b), the opposite is expected (E$_{VBM-Surf.}$ = valence-band maximum at the surface, E$_{VBM}$(x) = valence-band maximum in diamond).

Table 1:

| Diamond surface | Electron affinity $\chi$ | Dipole moment p |
|---|---|---|
| | Unit: *eV* | Unit: *e Å* |
| (111)-(2x1) | + 0.38 | |
| (111)-(2x1):H | - 1.27 | + 0.09 |
| (100)-(2x1) | + 0.50 | |
| (100)-(2x1):H | - 1.30 | + 0.08 |
| (100)-(1x1):O | + 1.73 | -0.10 |

Table I (data from Maier, F., Ristein, J., Ley, L. 2001; and Cui, J.B., Ristein, J., Ley, L., 1998).

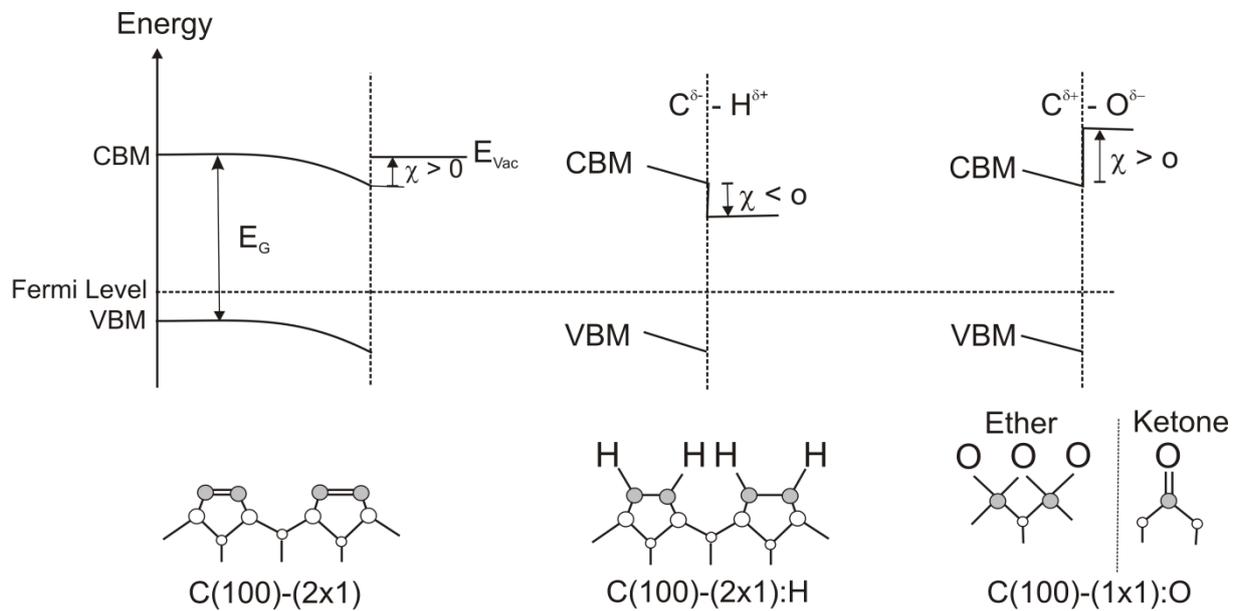

Fig. 1

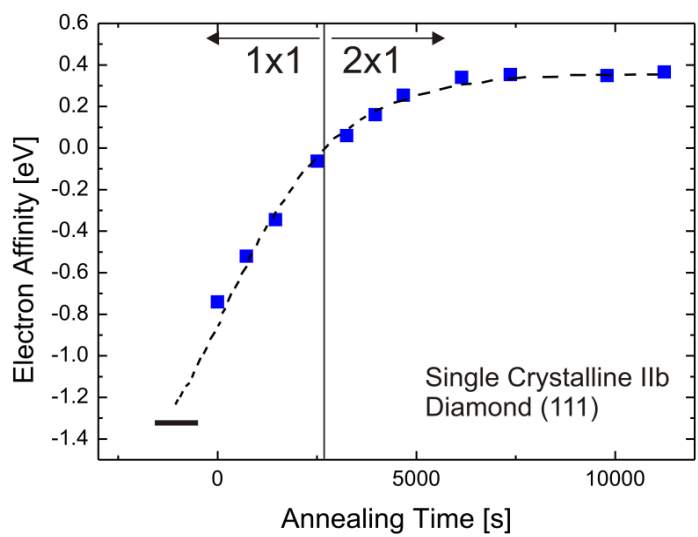

Fig. 2

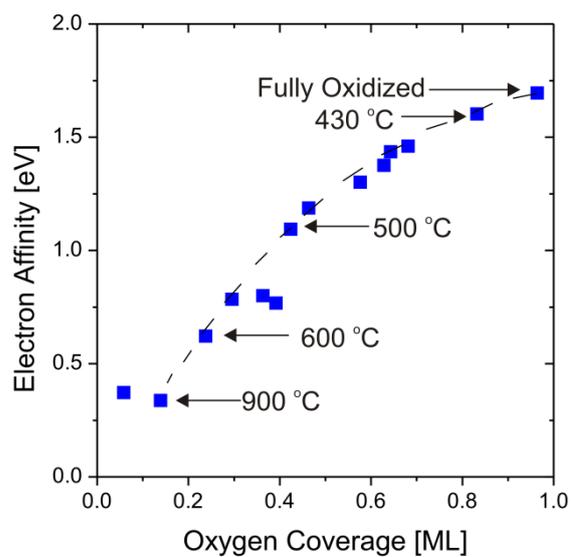

Fig. 3

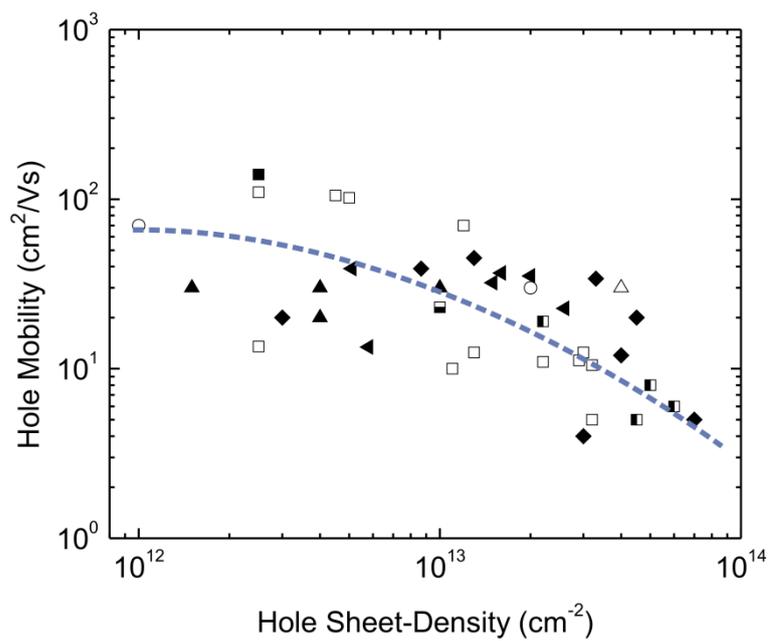

Fig. 4

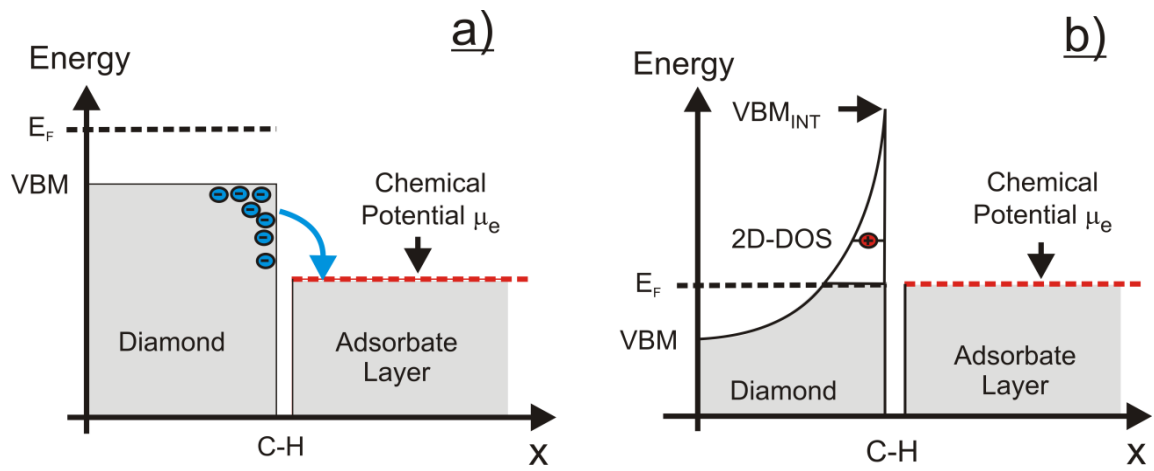

Fig. 5

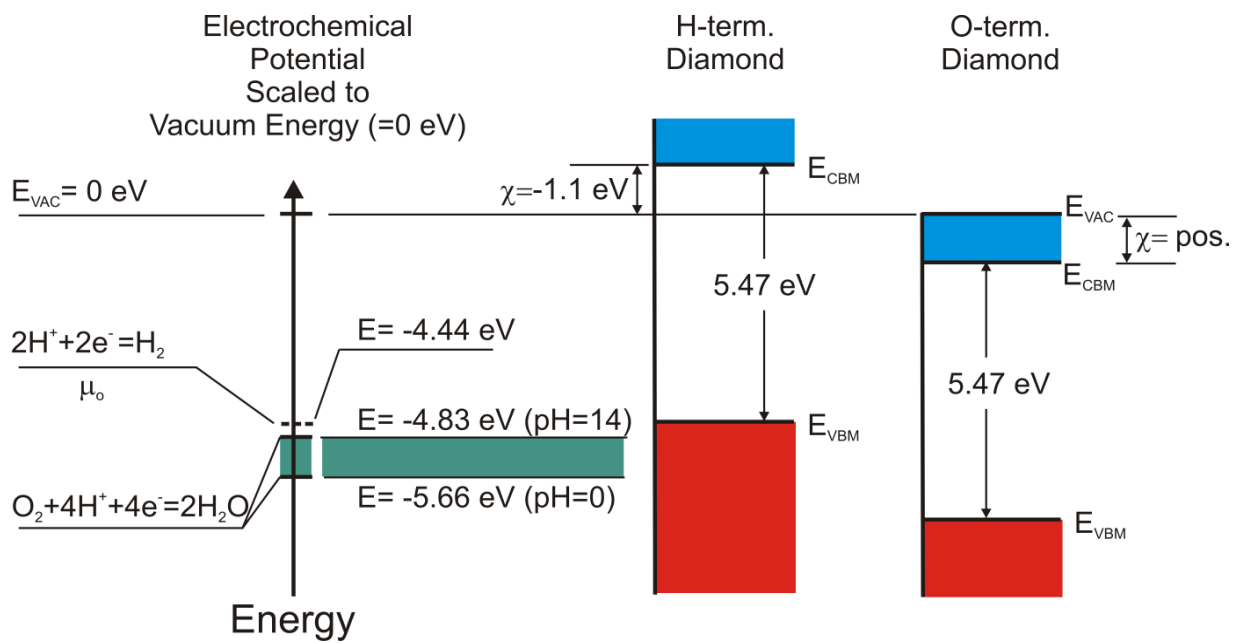

Fig. 6

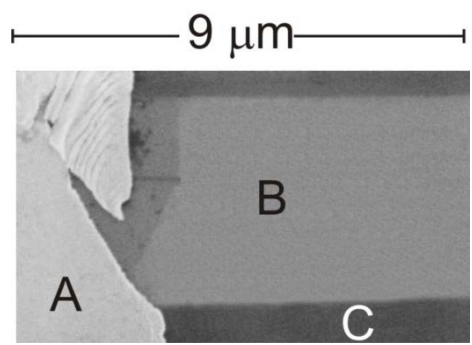

Fig. 7a

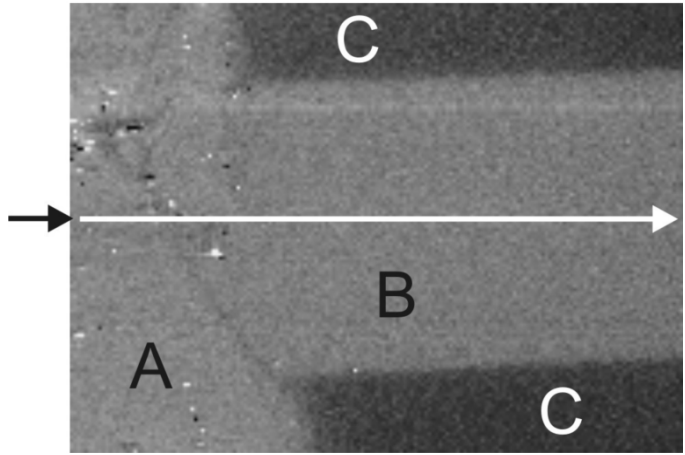

Fig. 7b

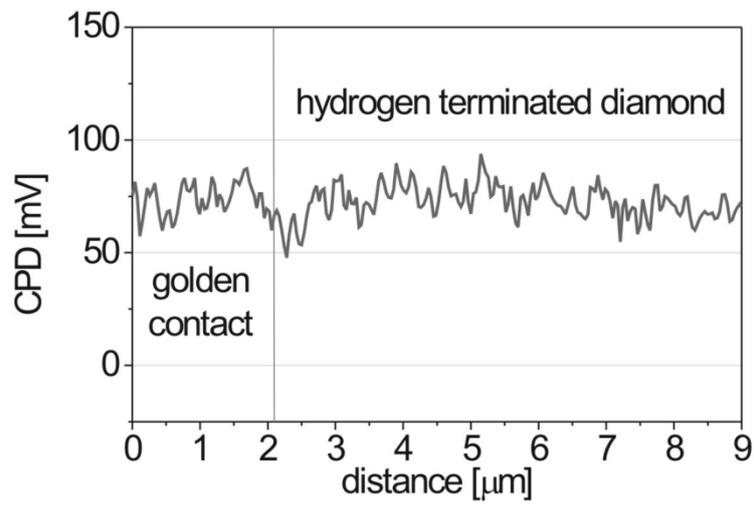

Fig. 7c

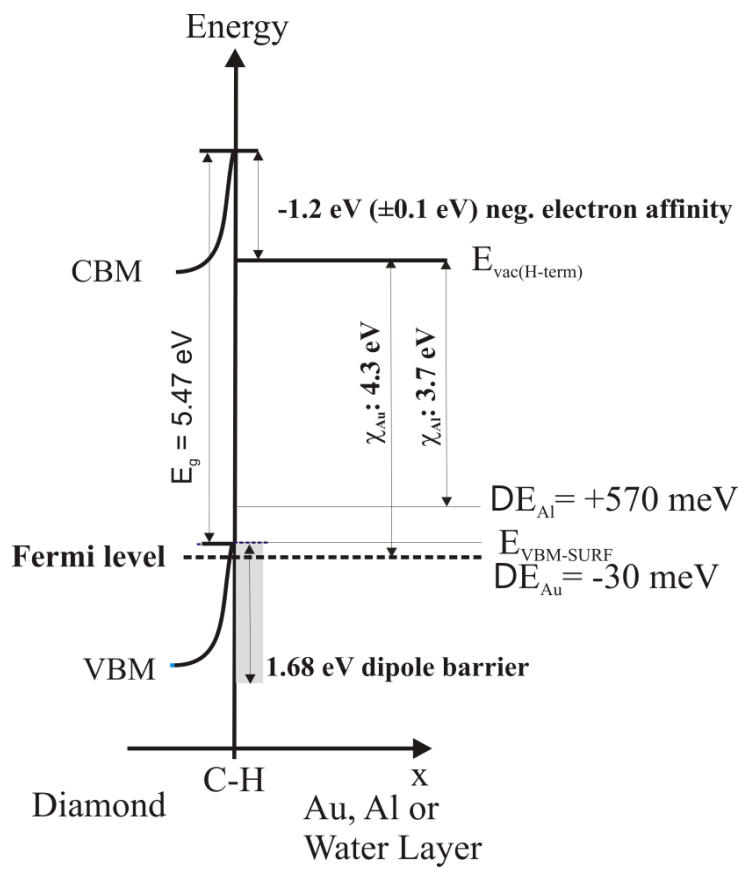

Fig. 8

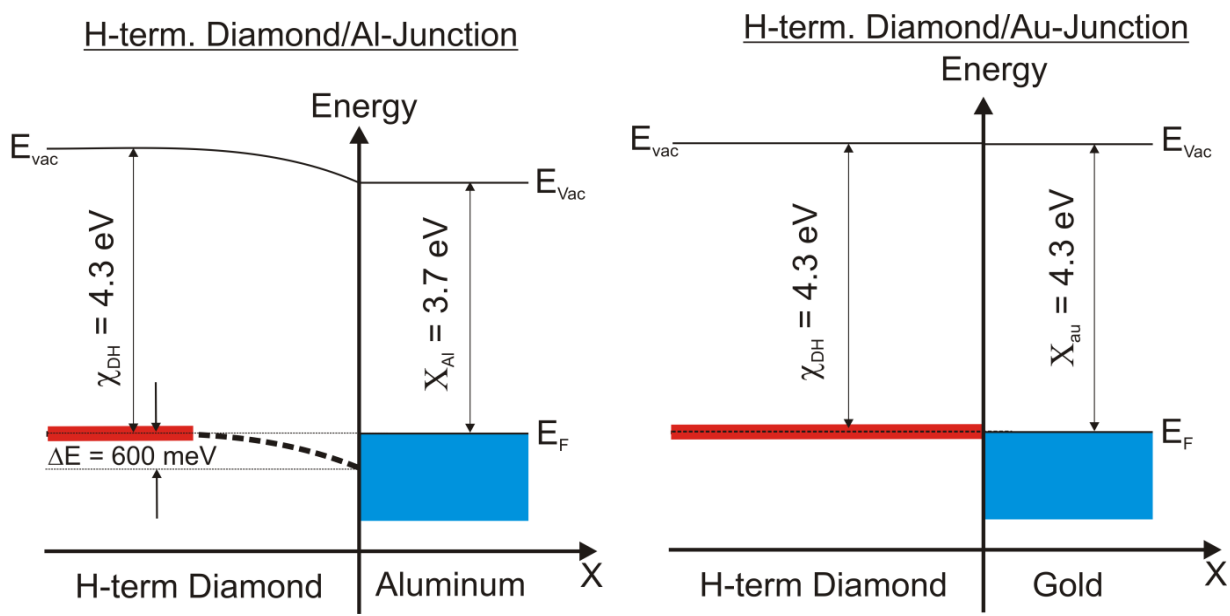

Fig. 9

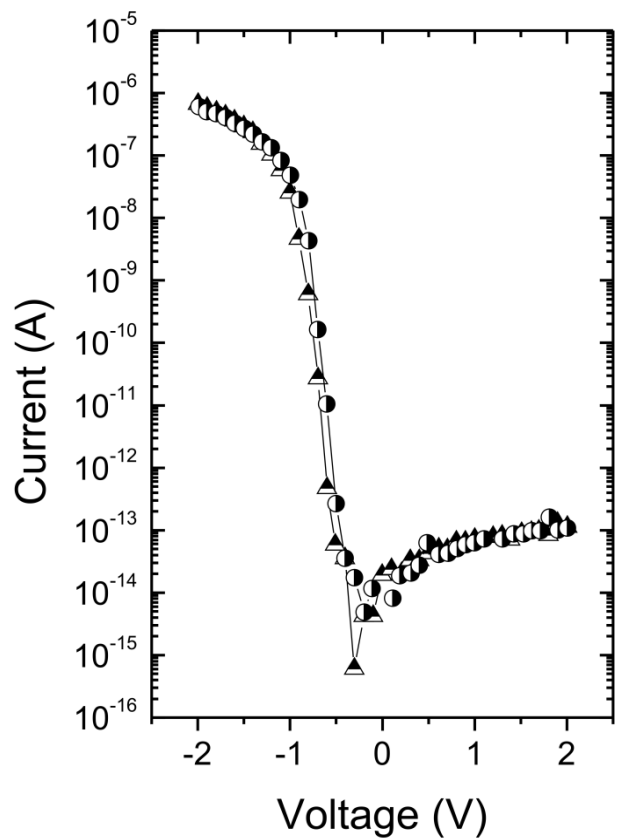

Fig. 10

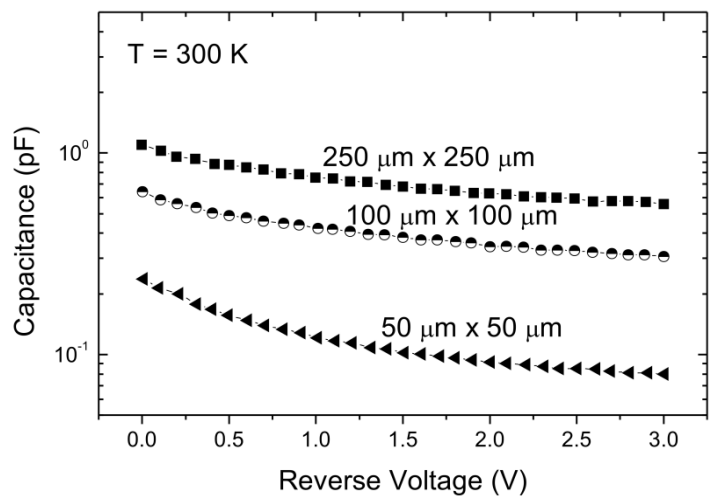

Fig. 11

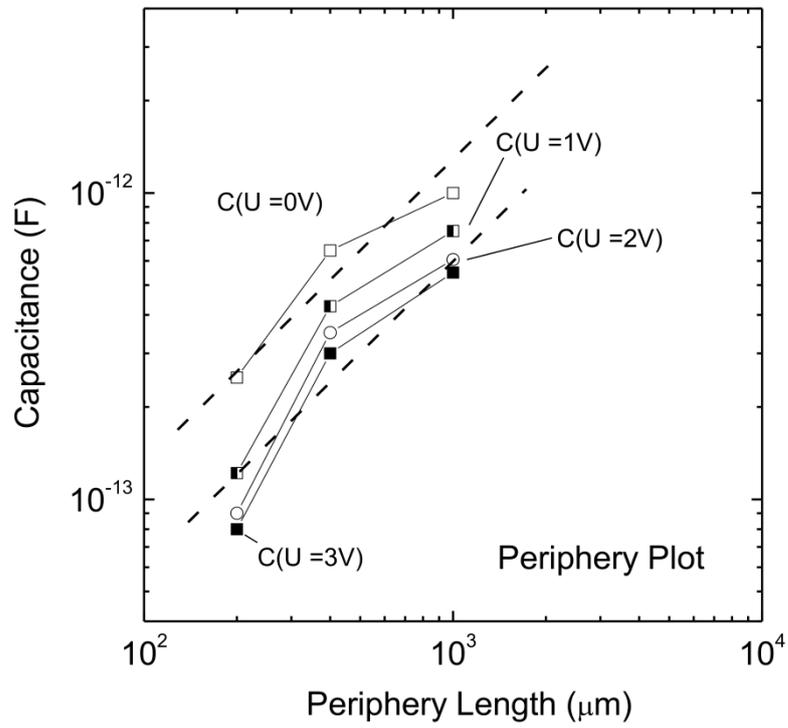

Fig. 12

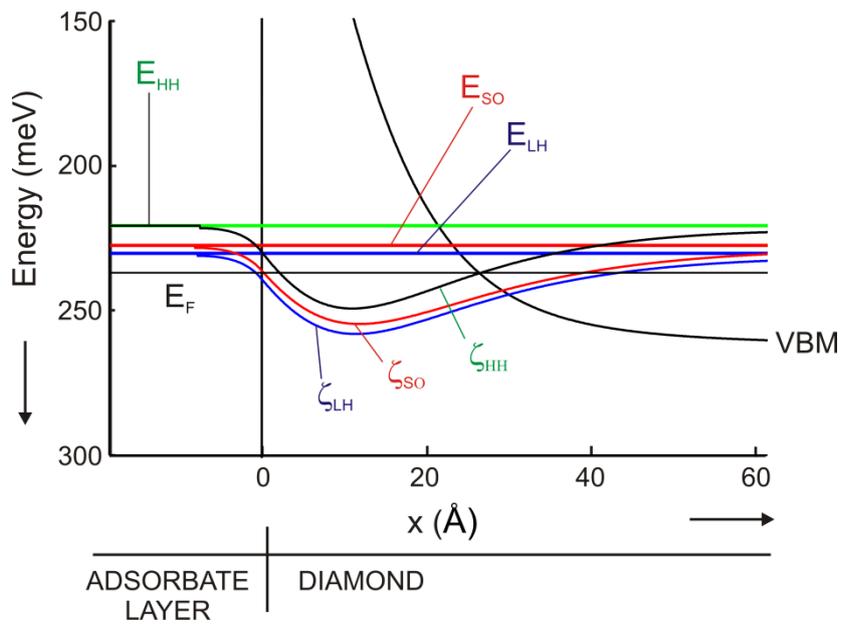

Fig. 13

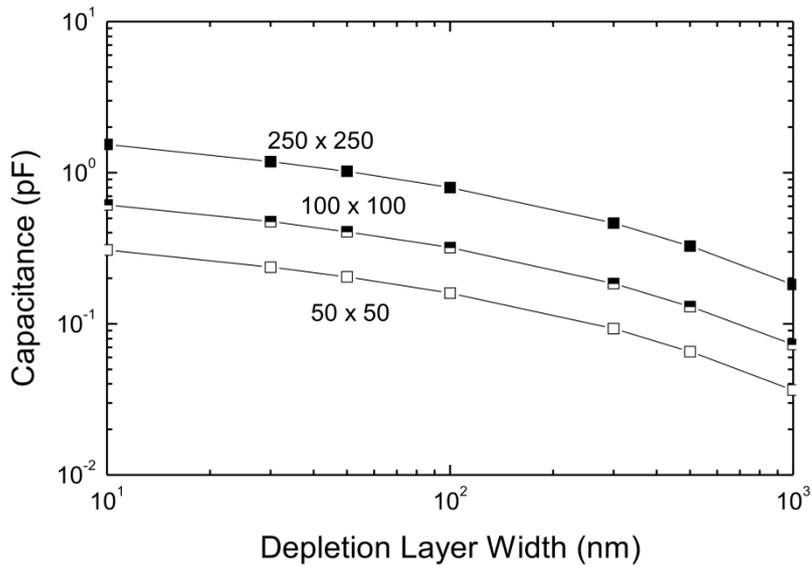

Fig. 14

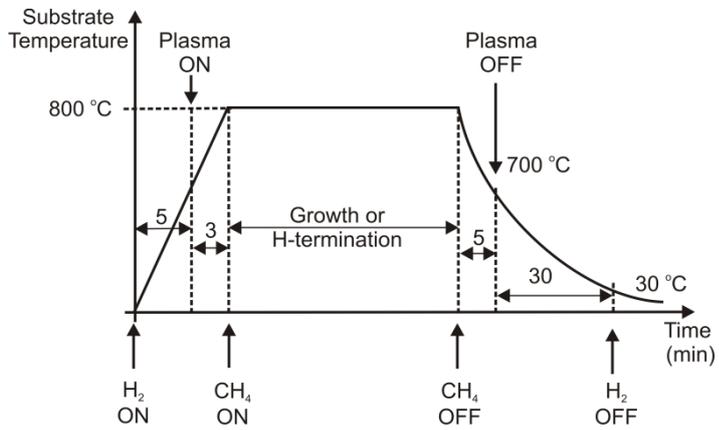

Fig. 15

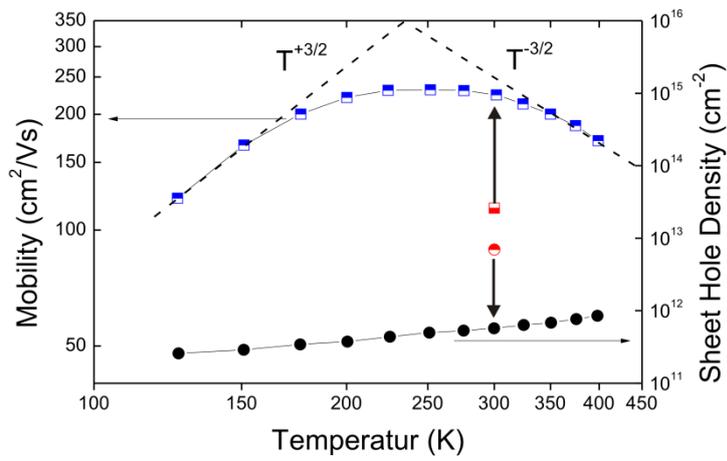

Fig. 16

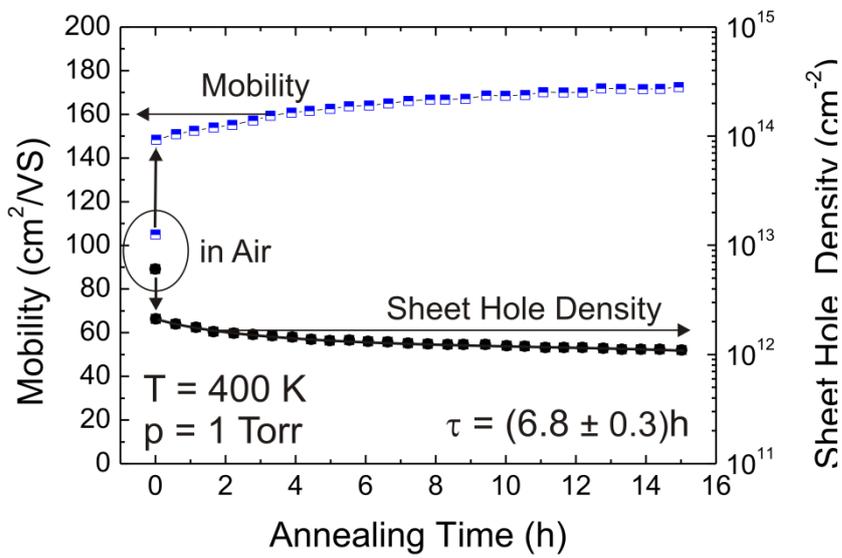

Fig. 17

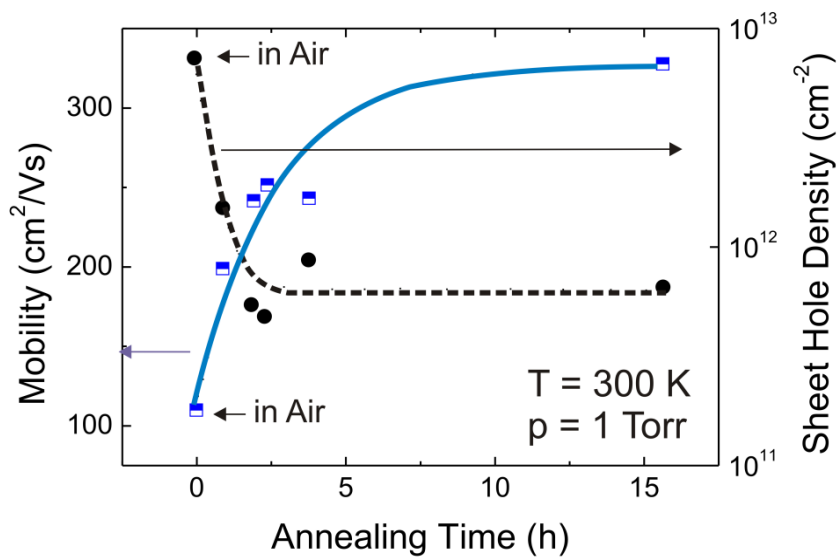

Fig. 18

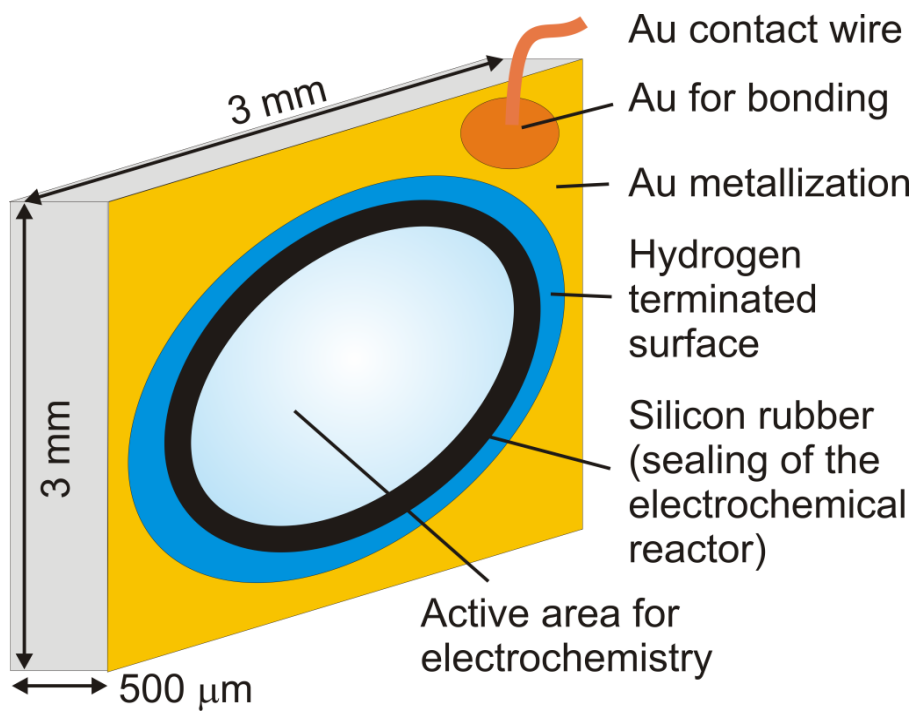

Fig. 19

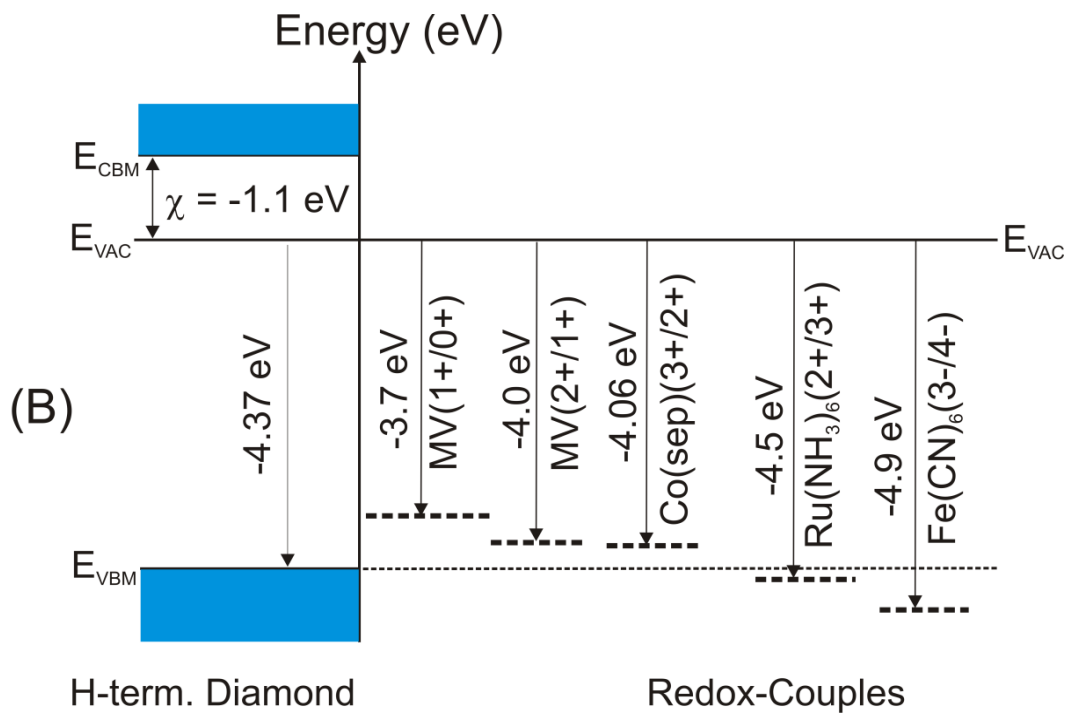

Fig. 20

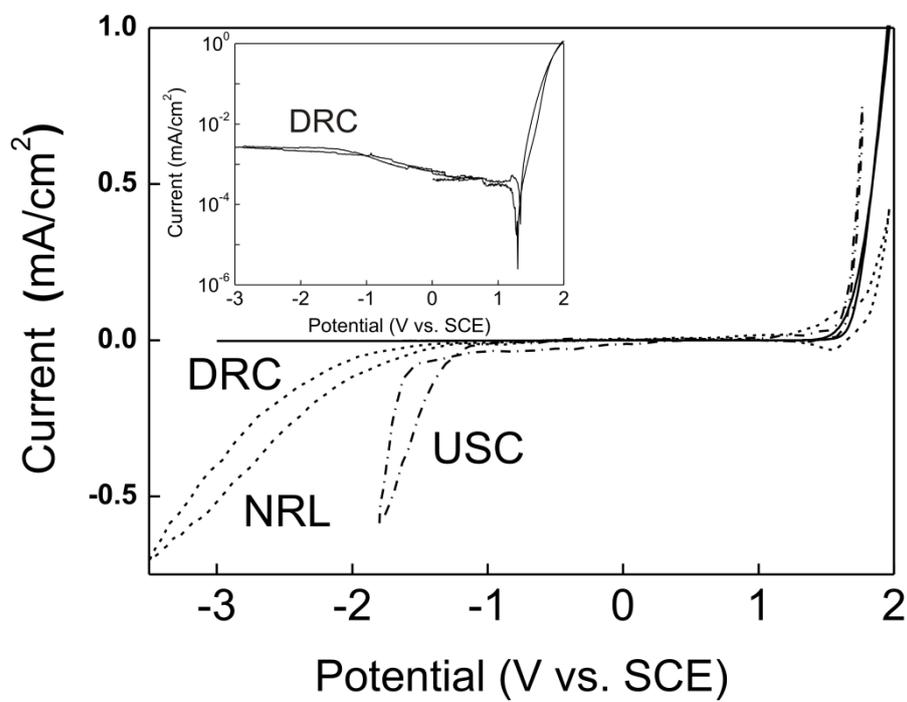

Fig. 21

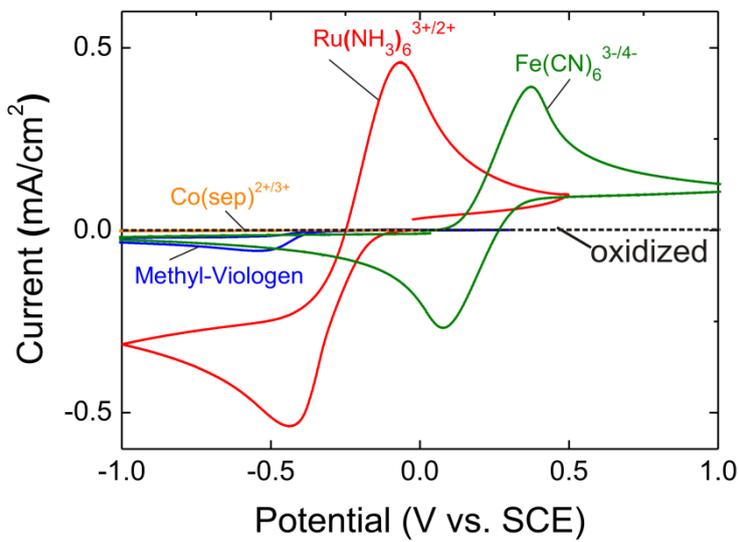

Fig. 22

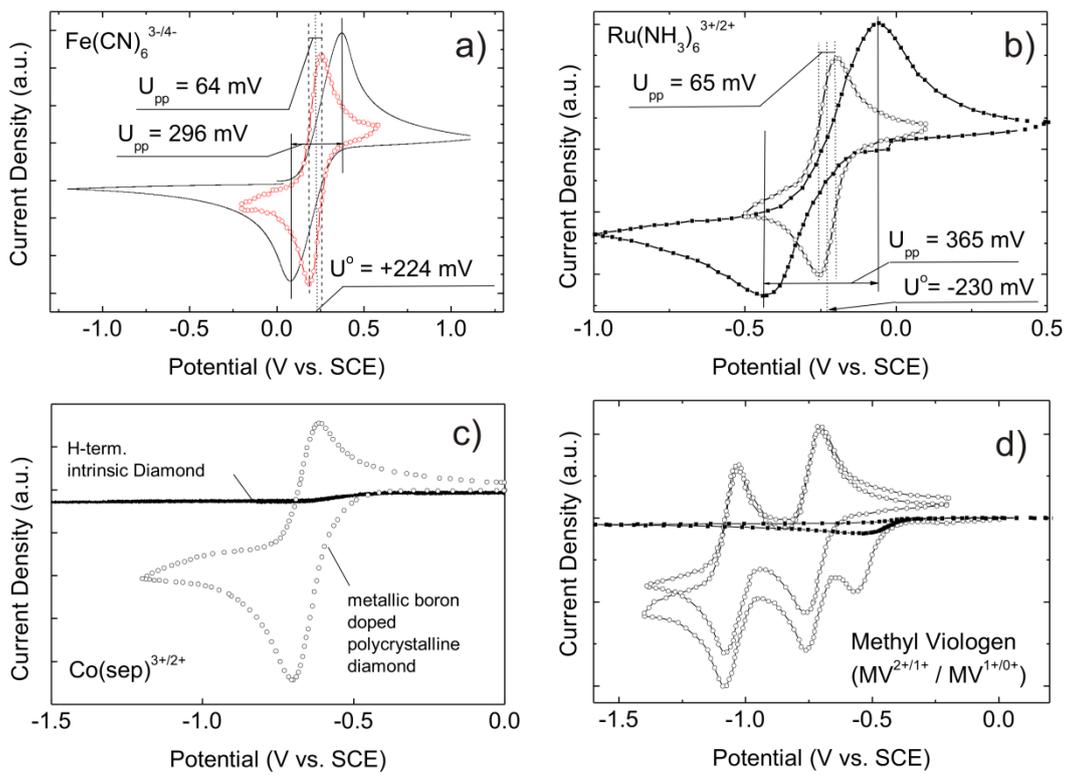

Fig. 23

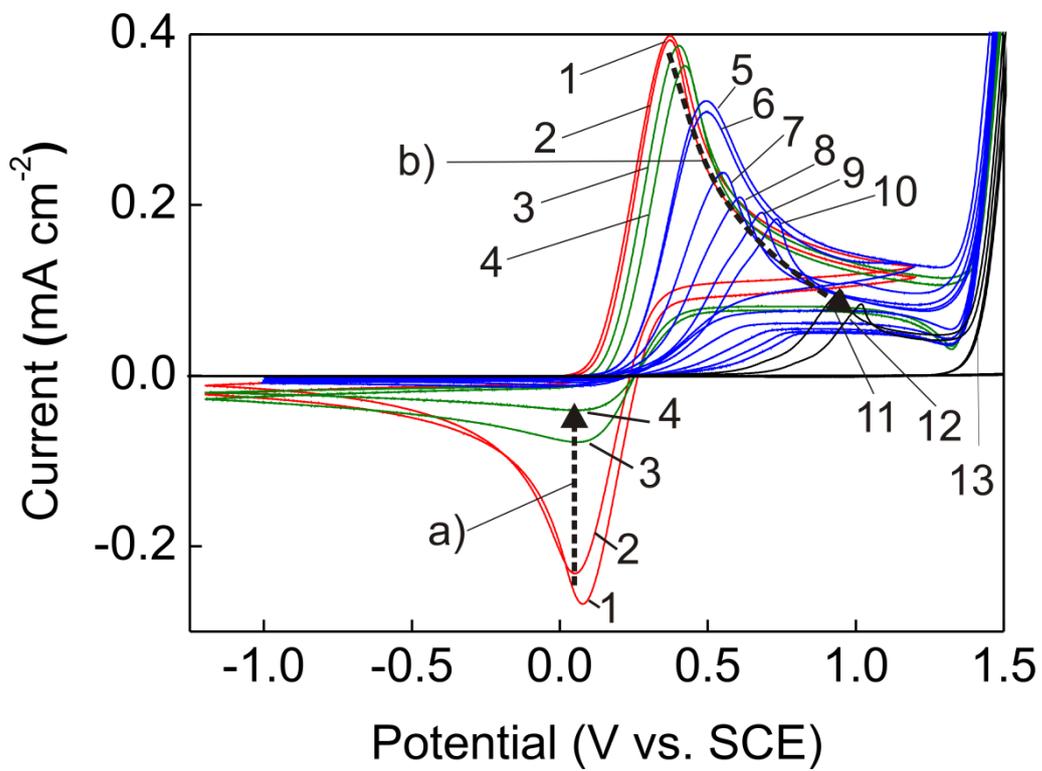

Fig. 24

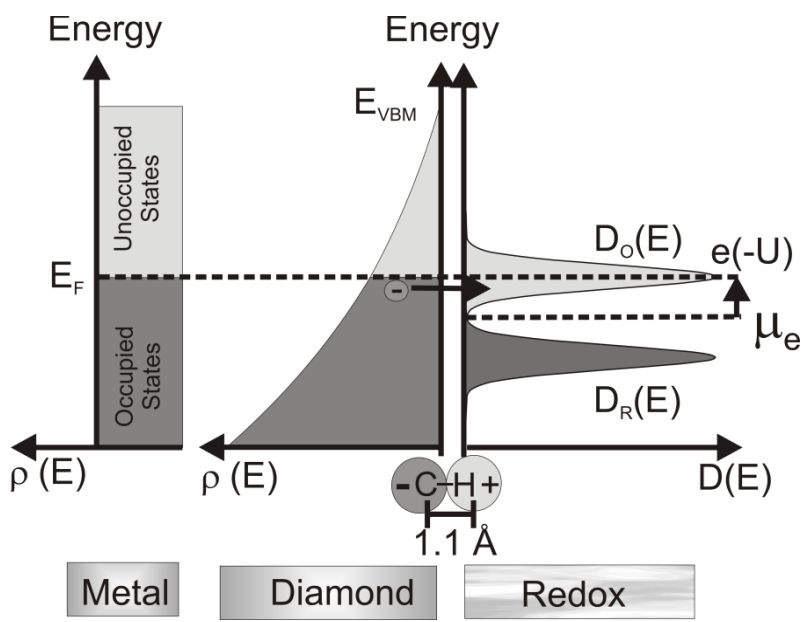

Fig. 25

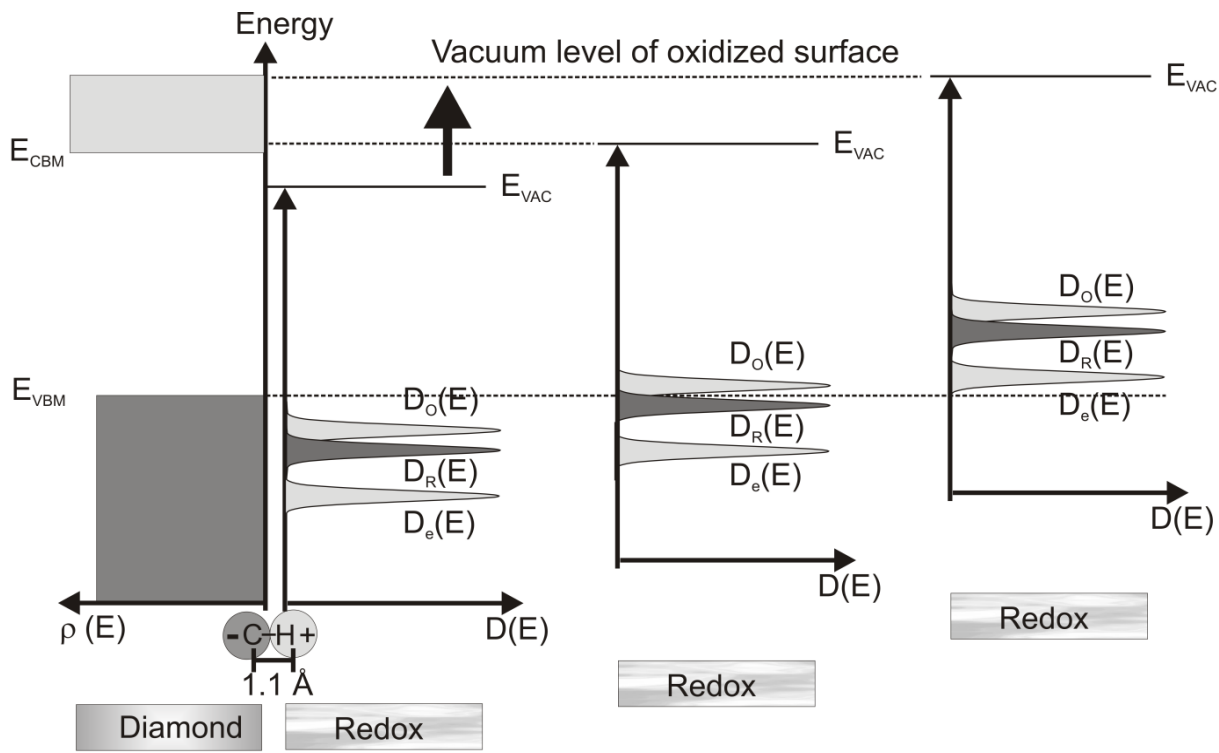

a) H-terminated    b) Partially Oxidized    c) Fully Oxidized

Fig. 26

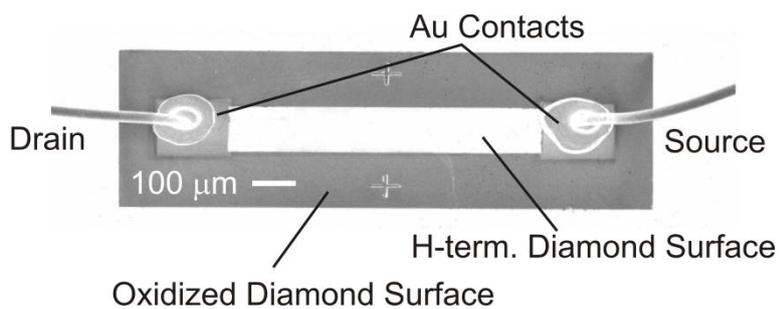

Fig. 27

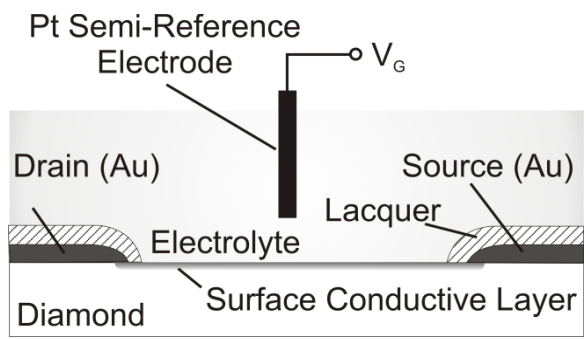

Fig. 28

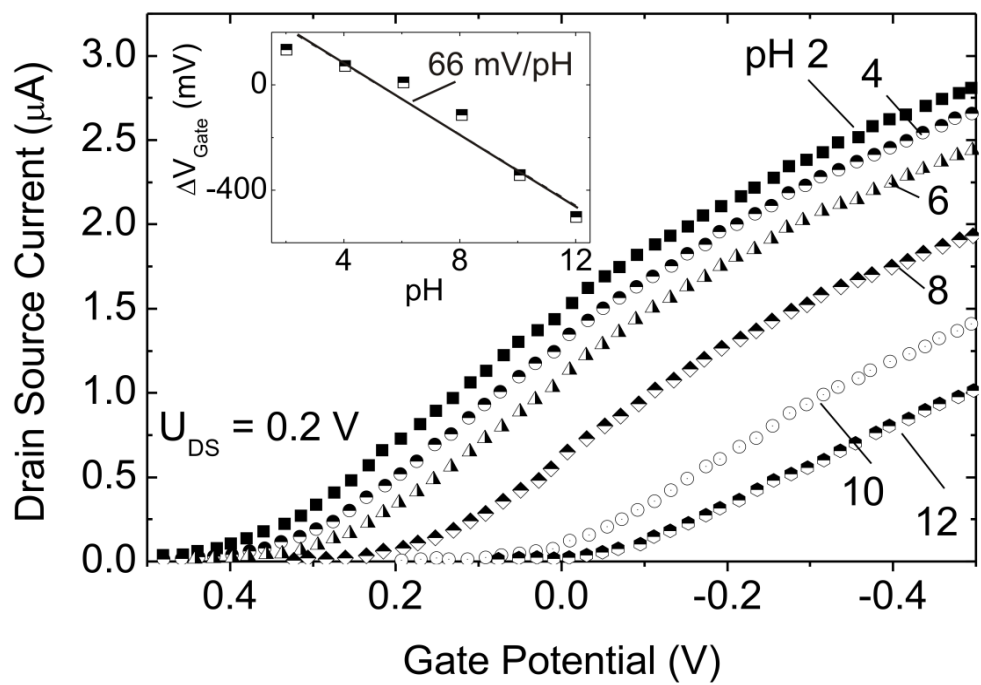

Fig. 29

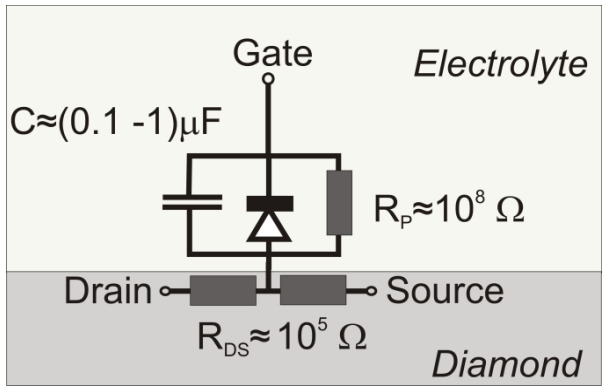

Fig. 30

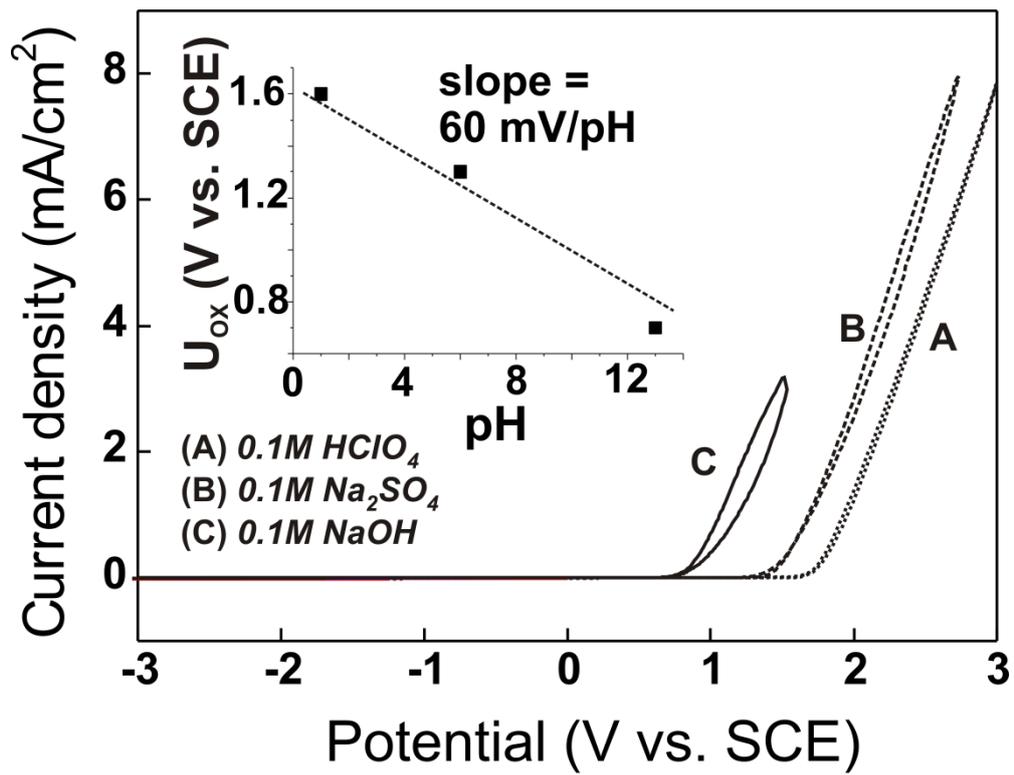

Fig. 31

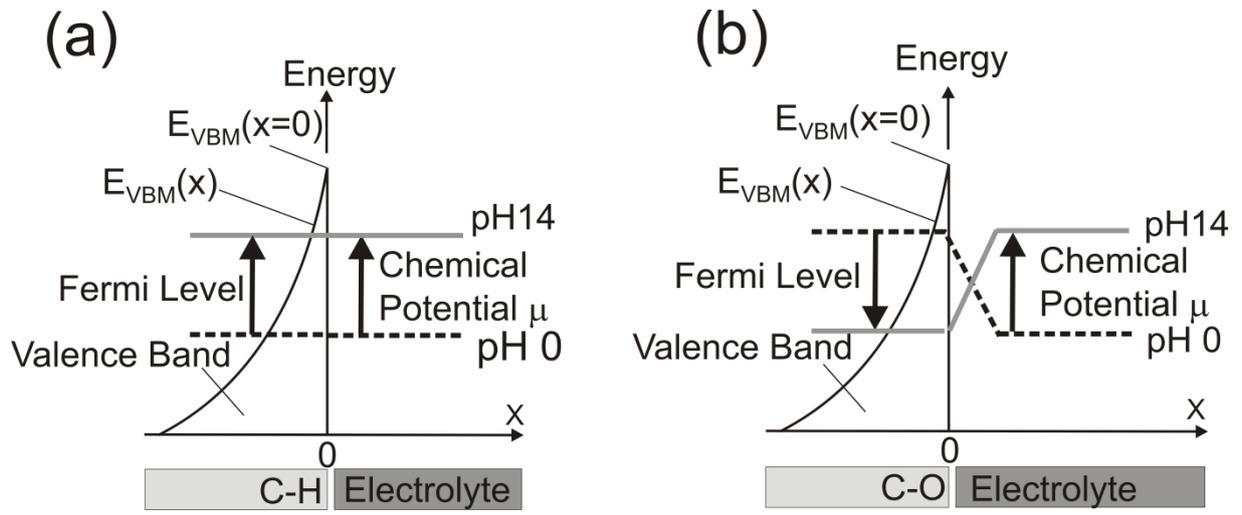

Fig. 32